\documentclass[aps,prb,twocolumn,showpacs,floatfix,twoside,superscriptaddress,eqsecnum]{revtex4-2}
\usepackage{amssymb}
\usepackage{graphicx}
\usepackage{amsmath}
\usepackage{amsfonts}
\usepackage{esint}
\usepackage{color}
\usepackage{xcolor}
\usepackage{float}
\usepackage{subcaption}
\usepackage{verbatim}
\usepackage[colorlinks=true,citecolor=blue,linkcolor=blue,urlcolor=blue]{hyperref}
\usepackage[justification=centerlast,font={small}]{caption}

\DeclareMathOperator*{\tr}{Tr}
\DeclareMathOperator{\im}{Im}
\DeclareMathOperator{\re}{Re}
\DeclareMathOperator{\sn}{sn}
\DeclareMathOperator{\sign}{sign}
\newcommand{\ttr}[1]{\tr{\left\{#1\right\}}}

\newcommand{\BT}{\mathcal{T}}

\newcommand{\BZ}{\mathcal{Z}}

\newcommand{\bp}{\mathbf{p}}
\newcommand{\bq}{\mathbf{q}}
\newcommand{\lint}{\int\limits}

\begin{document}

\title{Slowly decaying real-time oscillations in instanton crystal.}
\date{\today }
\author{Grigory A. Starkov}
\affiliation{Ruhr University Bochum, Faculty of Physics and Astronomy, Bochum, 44780,
Germany}
\email{Grigorii.Starkov@rub.de}
\author{Konstantin B. Efetov}
\affiliation{Ruhr University Bochum, Faculty of Physics and Astronomy, Bochum, 44780,
Germany}
\email{Konstantin.B.Efetov@rub.de}

\begin{abstract}
 Instanton crystal is a fascinating phase which is encountered when the minimum of the free energy corresponds to
a configuration with an imaginary-time-dependent order parameter in the form of a chain of alternating instantons
and anti-instantons.
We present the results of the investigation of the real-time correlation functions of the order parameter in the
instanton crystal phase. In order to obtain the correlation functions in real-time, we formulate an original method
of analytic continuation from imaginary times, which is easily adapted into an efficient numerical scheme for the
computations. The resulting correlation functions exhibit non-trivial slowly decaying oscillations in real-time, which is reminiscent of prethermal time crystals.

\end{abstract}

\maketitle

\section{Introduction.}

\subsection{Instanton crystal}

In standard theories of phase transitions one uses the concept of the order
parameter and spontaneous breaking of the symmetry in the phase when the
order parameter is not equal to zero \cite{landau}. As one considers
thermodynamic models starting with a time-independent Hamiltonian, physical
real time $t$ does not appear in thermodynamic quantities. In contrast, the
imaginary time $\tau $ is a usual additional \textquotedblleft
coordinate\textquotedblright\ that arises in the formalism of quantum field
theory which is nowadays a conventional tool for describing many body
systems.

In many cases, like spin models, the spin integration variables $\mathbf{S}%
\left( \tau \right) $ contain the single imaginary time variable $\tau .$ In
other cases, superconductivity, charge density waves, etc. the integration
variables are functions of two variables of the type $\Delta \left( \tau
,\tau ^{\prime }\right) $. As concerns the order parameter usually found
from the minimum of the free energy, it does not depend in all known cases
for spin systems on $\tau $ at all, while in the two-time cases it may
depend on the difference $\tau -\tau ^{\prime }.$ At the same time, it is
not clear whether a function of two variables $\tau $ and $\tau ^{\prime }$
playing the role of the order parameter may depend on $\tau +\tau ^{\prime
}, $ which would violate the imaginary time translation invariance. This
possibility does not contradict to general principles but one should be sure
that the free energy of such a state is minimal with respect to those of
other states.

The question about the possibility of the imaginary time-dependent order
parameter has been raised rather long ago \cite{mukhin,mukhin1}. In these
publications the order parameter was taken as a train of instantons and
anti-instantons but the free energy turned out to be higher than the one for
the static order parameter. More recently, the model was extended adding
additional interaction terms in publications \cite{mukhin2019,efetov2019}
but the free energies for the imaginary time-dependent states still exceeded
the free energy of the corresponding state with the static order parameter.
An attempt to construct soliton trains in imaginary time has been undertaken
in Ref. \cite{galitski}. In all these cases, however, the mathematical
construction of solitons hinges on the reflectionless potentials of the
underlying scattering problem. Generation of solitons using a supersymmetry
approach has been discussed in a recent paper \cite{galitski1}.
Apparently, in the models that can be solved exactly, solutions with the
imaginary time-dependent order parameter are energetically unfavorable and
one had to try more sophisticated models.

Fortunately, a proper model has been suggested in our recent publication
\cite{se}. This model contains the fermion part analogous to the one
considered in Refs. \cite{mukhin,mukhin1, mukhin2019,efetov2019} but the
fermions interact additionally with a boson mode. The crucial difference
with conventional electron-phonon models is that we include into
consideration the interaction of currents created by fermions and bosons
rather than interaction of densities. We have investigated the model in the
mean field approximation using both analytical and numerical methods and
demonstrated existence of the phase transition from the static phase into a
new state with the order parameter in a form of an instanton-antiinstanton
train in the imaginary time. The transition can be either of the first or of the
second order.

In~\cite{se}, we restricted ourselves to the calculation of thermodynamic quantities.
The discussion of the correlations of physical quantities in real time was left for the future study.
The reason for that was that we deemed this topic highly nont-trivial to deserve a separate work.


As such, this paper is devoted precisely to the study of real-time correlations in the instanton
crystal. We find an interesting behavior which is specific for this
thermodynamically stable state. 

The real time $t$ is
introduced via the standard quantum mechanical replacement of the operators
\begin{equation}
\hat{A}\left( t\right) =e^{it(\hat{H}-\mu\hat N)}\hat{A}e^{-it(\hat{H}-\mu\hat N)},  \label{a1}
\end{equation}%
where $\hat{H}$ is the full (time-independent) Hamiltonian of the model and $\mu$ is chemical potential.
Within this approach we calculate certain correlation functions assuming
that the system itself remains in the thermodynamic equilibrium.

The thermodynamic average $\left\langle ...\right\rangle $ describing these
functions equals
\begin{equation}
\left\langle ...\right\rangle =Z^{-1}Tr\left[ \left( ...\right) \exp \left[
-\beta (\hat{H}-\mu\hat{N})\right] \right] ,  \label{a3}
\end{equation}%
while
\begin{equation*}
Z=Tr\left[ e^{-\beta (\hat{H}-\mu\hat{N})}\right] ,
\end{equation*}%
is the partition function, and $\beta =1/T$ is the inverse temperature. The
brackets in Eq. (\ref{a3}) can contain products of real-time dependent
operators like those, given in Eq. (\ref{a1}).

So, we consider purely thermodynamic model but are going to obtain non-trivial
real-time dependent correlations. Of course, $\langle \hat{A}\left(
t\right) \rangle =\langle \hat{A}\rangle $ cannot depend on
$t$, and the perpetual motion is impossible.

\subsection{Instanton crystal vs quantum time crystal.}

Recently, a lot of attention has been paid to the study of time crystals in
several communities including condensed matter, atomic physics, quantum
optics. This activity was triggered by Wilczek \cite{wilczek} who proposed a
concept of quantum time crystals using a rather simple model that possessed
a state with a current oscillating in time. However, a more careful
consideration of the model \cite{bruno} has led to the conclusion that this
was not the equilibrium state. These publications were followed by
discussions of the possibility of realization of a thermodynamically stable
quantum time crystal \cite{wilczek1,li,bruno1,bruno2,nozieres,wilczek2}.

Meanwhile, it has been realized that long living oscillations can exist in
systems out of equilibrium \cite{volovik,sacha,sondhi1,sondhi2,nayak,yao}.
In particular, in the systems under periodic pumping from external source.
Several experimental works \cite{autti,zhang,choi} have nicely confirmed
that studying long-living or infinitely living time oscillations in such a context is a very interesting subject of
research as well. Actually, nowadays, the term \textquotedblleft quantum time
crystal\textquotedblright\ relates to this type of systems. A review on this
subject has appeared recently \cite{sondhi3}.

This shift of the interest from thermodynamic to non-equilibrium systems has
been greatly affected by the \textquotedblleft no-go\textquotedblright\
theorem proposed in Ref. \cite{watanabe}. According to the statements of
this theorem, the thermodynamic time crystals are not possible. The proof
looks mathematically convincing and, therefore,
the study of the equilibrium systems was discontinued naturally in favour of the non-equilibrium systems, while the name for this interesting subject of research stayed the same.


In this paper, we demonstrate that the real-time correlation functions of the Instanton Crystal can oscillate without significant decay for the prolonged periods of time. Unfortunately, the analysis of the current paper does not allow us to say conclusively if these oscillations survive in the asymptotic limit $t\rightarrow \infty$.
Overall,
the instanton crystal is a genuine thermodynamic state
obtained within a thermodynamic model using traditional methods of
theoretical physics. The oscillating character of the correlation functions
in real time is to some extent a \textquotedblleft
by-product\textquotedblright\ of the equilibrium properties of this state. This is the reason why we
avoid using the notion \textquotedblleft time crystal\textquotedblright\ for
the new phase found in our case.

Whether the observed oscillations are decaying or non-decaying, the sole fact of their appearance contradicts the statement of the \textquotedblleft no-go\textquotedblright\ theorem.
It turns out that, although their arguments are rather generally correct, the proof is not devoid of the problems. In particular, it was discussed in Ref.~\cite{sondhi3} that the proof in the case of non-zero temperature contains certain holes. We discuss the validity of the \textquotedblleft no-go\textquotedblright\ theorem in regards to the Instanton Crystal state in Sec.~\ref{section:validity_no_go}.


Finally, we should point out that we perform all the calculations in the paper using the mean-field approximation, which is a standard first stage in the analysis of a new phase transition.
We reserve the treatment of the effects of the fluctuations around the mean-field configurations for the future detailed study.

\subsection{The structure of the paper}

In Sec.~\ref{section:overview}, we formulate the model and recapitulate the mean field equations
derived previously \cite{se}.

In Sec.~\ref{section:continuation}, we discuss the analytical continuation of correlation functions from the imaginary-time axis to the real-time axis. We show that the standard method of continuation via frequency domain is ambiguous in the case of Instanton Crystal. Instead, we formulate a method of continuation directly in the time domain, which can be easily adapted into an efficient scheme for numerical computations.

In Sec.~\ref{section:analytics}, we consider the limit of negligible interaction between the fermions and the boson modes analytically.

In Sec.~\ref{section:numerics}, we present the results of the numerical investigation of the correlation function in real time
using an original scheme of computations.


In Sec.~\ref{section:validity_no_go} we discuss the validity of the \textquotedblleft
no-go\textquotedblright\ theorem of Ref. \cite{watanabe} in the case of Instanton Crystal State.

Section~\ref{section:discussion} contains discussion and concluding remarks.

In Appendix~\ref{section:scattering}, we discuss the connection between the correlation function studied in this paper and magnetic neutron scattering.

In Appendix~\ref{section:scheme}, we outline the details of the numerical scheme used to produce the results of Sec.~\ref{section:numerics}. The scheme is based on the method discussed in Sec.~\ref{section:continuation}.

\section{Instanton Crystal\label{section:overview}}

\subsection{Model\label{section:model}}

\begin{figure}[t]
\includegraphics[width=2.57in]{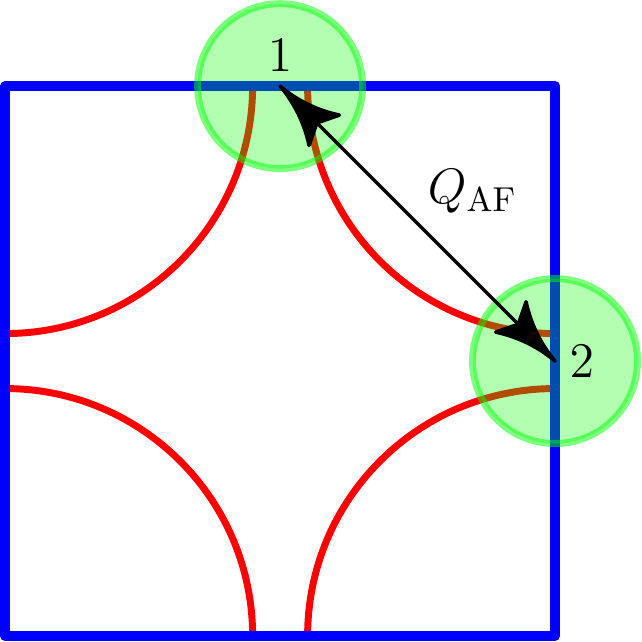}
\caption{Fermi surface (red) and overlapping hot spots (green).}
\label{fermi_surface}
\end{figure}

The model exhibiting the Instanton Crystal State was proposed in \cite{se}
on the basis of the Spin-Fermion model with Overlapping Hot Spots (SFMOHS)
studied previously in \cite{volkov1, volkov2, volkov3, efetov2019}.
In the Lagrangian formulation, the grand canonical partition function is given by the functional integral
\begin{multline}
\mathcal{Z}=\ttr{e^{-\beta(\hat{H}-\mu\hat{N})}}
=\\ =
\int \exp \left( -S\left[ \chi ,\chi ^{+},a\right] \right) D\chi
D\chi^{+}Da,  \label{zdef}
\end{multline}
where the action of the model has the structure
\begin{equation}
S\left[\chi ,\chi ^{+},a\right] =S_{\mathrm{0}}+S_{\mathrm{int}}+S_{\mathrm{B%
}}+S_{\mathrm{FB}}.  \label{adef:full}
\end{equation}%
In Eq.~\eqref{adef:full}, the term $S_0$ stands for the action of a system of non-interacting fermions:
\begin{multline}
S_{\mathrm{0}}\left[ \chi ,\chi ^{+}\right] = \\
\* = \sum_{\mathbf{p}}\int\limits_{0}^{\beta}\chi _{\mathbf{p}}^{+}\left(
\tau \right) \left[ \left(\partial_{\tau}+\varepsilon^{+}_\mathbf{p}-\mu\right)%
\check{\mathrm{I}}+\varepsilon^{-}_\mathbf{p} \check{\Sigma}_{3}\right] \chi
_{\mathbf{p}}\left( \tau \right) d\tau.  \label{adef:fermion}
\end{multline}
These fermions live in two bands $1$ and $2$, which correspond to the two
anti-nodal "hot regions" of SFMOHS (see Fig.\ref{fermi_surface} for the
schematic representation of Fermi surface with hot regions).
Four-component vectors
\begin{equation}
 \chi_\bp^{(+)}(\tau)=\left(\chi_{1\bp}^{1(+)}(\tau),\chi_{1\bp}^{2(+)}(\tau),\chi_{2\bp}^{1(+)}(\tau),\chi_{2\bp}^{2(+)}(\tau)\right)
\end{equation}
contain as components Grasmann fields $\chi_{\alpha\bp}^{s(+)}(\tau)$ that correspond to destruction (creation) operators for the fermions from the bands $s=1,2$ with spin projection labeled by $\alpha = 1,2$.
The energies $%
\varepsilon _{\mathbf{p}}^{\pm }$ are expressed in terms of the spectra $%
\varepsilon _{1,2}\left( \mathbf{p}\right) $ in the bands $1,2$ as
\begin{equation}
\varepsilon _{\mathbf{p}}^{\pm }=\frac{1}{2}\left( \varepsilon _{1}\left(
\mathbf{p}\right) \pm \varepsilon _{2}\left( \mathbf{p}\right) \right) .
\label{a4}
\end{equation}%
The operators $\check{\Sigma}_{i}$, $i=1,2,3$ are Pauli matrices acting in
the subspace of the bands $1$ and $2$, while $\check{\mathrm{I}}$ is the
identity operator acting in the same subspace.

As usual, the imaginary time $\tau $ is defined for $0\leq \tau \leq \beta
\equiv 1/T$ where $T$ is the temperature. The fermionic fields $\chi
_{\alpha \mathbf{p}}^{s}\left( \tau \right) $, $\chi _{\alpha \mathbf{p}%
}^{s+}\left( \tau \right) $ obey standard antiperiodic boundary conditions
\begin{equation}
\chi _{\alpha \mathbf{p}}^{s}\left( \tau +\beta \right) =-\chi _{\alpha
\mathbf{p}}^{s}\left( \tau \right) ,\;\chi _{\alpha \mathbf{p}}^{s+}\left(
\tau +\beta \right) =-\chi _{\alpha \mathbf{p}}^{s+}\left( \tau \right) ,
\label{a17}
\end{equation}%

The second term $S_{\mathrm{int}}$ in Eq.~\eqref{adef:full} stands for
the interaction between the fermions from different bands
\begin{multline}
S_{\mathrm{int}}\left[ \chi ,\chi ^{+}\right] =-\cfrac{U_0}{4V}\sum_{\mathbf{%
p}_{1,}\mathbf{p}_{2},\mathbf{q}}\int\limits_{0}^{\beta }d\tau \left( \chi _{%
\mathbf{p}_{1}}^{+}\left( \tau \right) \check{\Sigma}_{2}\chi _{\mathbf{p}%
_{1}+\mathbf{q}}\left( \tau \right) \right) \times \\
\*\times \left( \chi _{\mathbf{p}_{2}}^{+}\left( \tau \right) \check{\Sigma}%
_{2}\chi _{\mathbf{p}_{2}-\mathbf{q}}\left( \tau \right) \right).
\label{adef:int}
\end{multline}%
where $V$ is the volume of the system. Physically, this term corresponds to
the local in time attraction of the fermionic loop currents $A_\bq(\tau)$
\begin{equation}
A_\bq(\tau)=\cfrac{1}{\sqrt{V}}\sum_{\mathbf{p}}\chi_{\mathbf{p}%
}^{+}\check{\Sigma}_{2}\chi_{\mathbf{p}+\mathbf{q}}(\tau).  \label{a8}
\end{equation}

The first two terms in Eq.~\eqref{adef:full} originate from SFMOHS. The
last two ones, on other hand, were specifically introduced in \cite{se} to
stabilize the instanton crystal state. The third one, $S_\mathrm{B}$, describes a system of current-like
bosonic modes labeled by different momenta $\mathbf{q}$:
\begin{equation}
S_{\mathrm{B}}\left[ a\right] =\frac{1}{U_{\mathrm{2}}}\sum_{\mathbf{q}%
}\int\limits_{0}^{\beta }\left[ \left\vert \frac{da_{\mathbf{q}}\left( \tau
\right) }{d\tau }\right\vert ^{2}+\omega _{\mathbf{q}}^{2}\left\vert a_{%
\mathbf{q}}\left( \tau \right) \right\vert ^{2}\right] d\tau .
\label{adef:boson}
\end{equation}%
where $a_{\mathbf{q}}(\tau )$ are periodic complex fields satisfying
\begin{equation}
(a_{\mathbf{q}}(\tau ))^{\ast }=a_{-\mathbf{q}}(\tau ).
\label{real_constraint}
\end{equation}%
The fields $a_{\mathbf{q}}\left( \tau \right) $
correspond to the coordinates in the language of oscillator modes, and $da_{%
\mathbf{q}}\left( \tau \right) /d\tau $ correspond to their velocities.

Finally, the last term, $S_\mathrm{FB}$, describes the coupling between the current-like bosonic modes and the fermions:
\begin{equation}
S_{\mathrm{FB}}\left[ \chi ,\chi ^{+},a\right] =-\sum_{\mathbf{p,q}}\int\limits_{0}^{\beta }A_\bq(\tau)  \frac{da_{\mathbf{q}}\left( \tau \right) }{%
d\tau }d\tau .  \label{adef:fb}
\end{equation}
The distinguishing feature of the current-like modes is that, instead of the
charge of the fermions, they couple to the fermionic loop currents $A_\bq(\tau)$.

Starting with Eq.~\eqref{zdef}, the current-like
modes can be integrated out leading to the appearance of the effective
non-local repulsion between the fermionic loop currents in addition to the
local attraction due to the term~\eqref{adef:int}. Then, the overall loop current
interaction can be decoupled by the means of Hubbard-Stratonovich
transformation. Finally, one can integrate out the fermions to obtain an
effective description of the partition function $\mathcal{Z}$ in terms of
the functional integral over the real order parameter field $b(\tau )$. If
we neglect the spatial fluctuations, we can write it as
\begin{equation}
\mathcal{Z}=\int \mathcal{D}[b(\tau )]\exp {\left( -\beta \mathcal{F}%
[b]\right) },  \label{landau_path_integral}
\end{equation}%
where the free energy functional is
\begin{multline}
\frac{\beta \mathcal{F}[b]}{V}=\iint\limits_{0}^{\beta }d\tau d\tau ^{\prime
}K^{-1}(\tau -\tau ^{\prime }|\omega _{0})b(\tau )b(\tau ^{\prime })- \\
\*-2\int \frac{d\mathbf{p}}{(2\pi )^{2}}\tr_{s,\tau }\left[ \ln {\left( %
\vphantom{\int}(\partial _{\tau }+\varepsilon _{\mathbf{p}}^{+}-\mu)\check{%
\mathrm{I}}+\varepsilon _{\mathbf{p}}^{-}\check{\Sigma}_{3}-b(\tau )\check{%
\Sigma}_{2}\right) }\right] .\label{fe_functional}
\end{multline}%
%
%
%
%
%
%
%
%
%
%
%
%
%
%
%
%
%
%
%
%
Non-local kernel $K^{-1}(\tau -\tau ^{\prime }|\omega _{0})$ is given by
\begin{multline}
K^{-1}(\tau -\tau ^{\prime }|\omega _{0})= \\
\*=\frac{1}{U_{0}+U_{2}}\left[ \delta (\tau -\tau ^{\prime })+\frac{U_{2}}{%
U_{0}}K_{0}(\tau -\tau ^{\prime }|\tilde{\omega}_{0})\right] ,
\label{kernel_inverse}
\end{multline}%
where
\begin{equation}
K_{0}(\tau -\tau ^{\prime }|\tilde{\omega}_{0})=\frac{\tilde{\omega}%
_{0}\cosh {\left[ \tilde{\omega}_{0}\left( \frac{\beta }{2}-|\tau -\tau
^{\prime }|\right) \right] }}{2\sinh {\frac{\beta \tilde{\omega}_{0}}{2}}},
\end{equation}%
and
\begin{equation}
\tilde{\omega}_{0}=\sqrt{\frac{U_{0}}{U_{0}+U_{2}}}\omega _{0}
\end{equation}%
is the renormalized frequency of the current-like mode.

The
order parameter field $b(\tau)$ satisfies the periodic boundary conditions:
\begin{equation}
b(\tau) = b(\tau + \beta).
\end{equation}

\subsection{Mean-Field Analysis}

Mean-field solutions correspond to the stationary configurations of the free
energy functional $\mathcal{F}[b(\tau)]$:
\begin{multline}
b(\tau) + \frac{U_2}{U_0} \int\limits_0^\beta d\tau^\prime
K_0(\tau-\tau^\prime|\tilde\omega_0) b(\tau^\prime) = \\
\* = -(U_0+U_2)\int\frac{d\mathbf{p}}{(2\pi)^d}\tr{\left\{\check\Sigma_2%
\check G(\tau,\tau)\right\}},  \label{gap:general}
\end{multline}
where the fermion Green's function $\check G(\tau,\tau^\prime)$ satisfies
\begin{multline}
\left[(\partial_\tau + \varepsilon_\mathbf{p}^+)\check{\mathrm{I}} +
\varepsilon_\mathbf{p}^-\check\Sigma_3 - b(\tau)\check\Sigma_2\right]\check
G(\tau,\tau^\prime) = \\
\* = -\check{\mathrm{I}}\delta(\tau-\tau^\prime).  \label{greens_function}
\end{multline}

In the absence of non-local interaction, i.e., for $U_{2}=0$, the solutions
of Eq.~\eqref{gap:general} are known analytically. The minimum of the free
energy functional corresponds to the stationary configurations $b(\tau
)\equiv \pm \gamma _{T}$, where the static gap $\gamma _{T}$ is determined
from the self-consistency equation
\begin{multline}
\cfrac{2}{U_0}= \\
\*=\int \cfrac{d\mathbf{p}}{(2\pi)^2}\cfrac{\tanh{\frac{\beta(\kappa_%
\mathbf{p}^{(0)} + \varepsilon^+_\mathbf{p})}{2}} +
\tanh{\frac{\beta(\kappa_\mathbf{p}^{(0)} -
\varepsilon^+_\mathbf{p})}{2}}}{\kappa_\mathbf{p}^{(0)}}.
\label{self_consistency:static}
\end{multline}%
with
\begin{equation}
\kappa _{\mathbf{p}}^{(0)}=\sqrt{(\varepsilon _{\mathbf{p}}^{-})^{2}+\gamma
_{T}^{2}}
\end{equation}%
Besides that, there is also a series of saddle point solutions which
correspond to the alternating chains of instantons and anti-instantons \cite%
{mukhin,mukhin1,mukhin2,mukhin2019,efetov2019,se}. These
instanton-antiinstanton solutions are expressed exactly in terms of Jacobi
elliptic function $sn(x|k)$:
\begin{equation}
b(\tau )=k\gamma \sn(\gamma (\tau -\tau _{0})|k).\label{jacobi_order_parameter}
\end{equation}%
The period of instanton-antiinstanton pair is
\begin{equation}
W=4K(k)/\gamma  \label{p0}
\end{equation}%
where $K(k)$ is the complete elliptic integral of the first kind. For a
solution with $m$ instanton-antiinstanton pairs, the requirement of
periodicity leads to the condition
\begin{equation}
\beta =m\times \frac{4K(k)}{\gamma }.  \label{iai_periodicity}
\end{equation}%
In order to determine parameters $k$ and $\gamma $ unambiguously, Eq.~%
\eqref{iai_periodicity} should be supplemented by generalized
self-consistency equation
\begin{multline}
\cfrac{2}{U_0}= \\
\*=\int \cfrac{d\mathbf{p}}{(2\pi)^2}\cfrac{|\varepsilon^-_\mathbf{p}|\left[%
\tanh{\frac{\beta(\kappa_\mathbf{p} + \varepsilon^+)}{2} +
\tanh{\frac{\beta(\kappa_\mathbf{p} - \varepsilon^+)}{2}}} \right]}
{\sqrt{\left((\varepsilon_\mathbf{p}^-)^2 +
\gamma^2\frac{(1-k)^2}{4}\right)\left((\varepsilon_\mathbf{p}^-)^2 +
\gamma^2\frac{(1+k)^2}{4}\right)}},  \label{self_consistency}
\end{multline}%
where
\begin{equation}
\kappa _{\mathbf{p}}=|\varepsilon _{\mathbf{p}}^{-}|\sqrt{%
\cfrac{\left((\varepsilon_\mathbf{p}^-)^2 +
\gamma^2\frac{(1-k)^2}{4}\right)}{\left((\varepsilon_\mathbf{p}^-)^2 +
\gamma^2\frac{(1+k)^2}{4}\right)}}\cfrac{\Pi(n, \tilde{k})}{K(\tilde{k})},
\label{kappa}
\end{equation}%
where
\begin{equation}
n=\cfrac{\gamma^2k}{(\varepsilon_\mathbf{p}^-)^2 + \gamma^2\frac{(1+k)^2}{4}}%
,\quad \tilde{k}=\cfrac{2\sqrt{k}}{1+k}.  \label{n_ktilde}
\end{equation}%
Here, $\Pi (n,k)$ is the complete elliptic integral of the third kind. (One
can find more details about elliptic integrals and functions, for example,
in \cite{whittaker,as}). We should also point out that in the limit $%
k\rightarrow 1$, which corresponds to the dilute instanton-antiinstanton
approximation, Eq.~\eqref{self_consistency} reduces to Eq.~%
\eqref{self_consistency:static}.

As we have already mentioned, in the absence of non-local interaction, the
minimum of the free energy functional corresponds to static order parameter
field. However, when $U_{\mathrm{2}}$ is turned on, the energy of the static
configuration can get pushed higher than the energy of instanton
configurations. In this case, the minimum corresponds to a solution with a
particular number $m$ of instanton-antiinstanton pairs. In~Ref. \cite{se},
we demonstrated this phenomena both analytically (for $U_{\mathrm{2}}/U_{%
\mathrm{0}}\ll 1$) and numerically. In the latter case, we used an efficient
procedure based on direct minimization of suitably discretized version of
the free energy functional.

\section{Analytic Continuation of the order parameter correlation function.\label{section:continuation}}

\subsection{Real- and imaginary- time correlation functions.\label{section:cont-introduction}}

The experimentally relevant quantity important for the characterization of
the Instanton Crystal State is the dynamic correlation function of the loop
currents:
\begin{equation}
C^{\mathcal{T}}(t)=\frac{1}{V\mathcal{Z}}\tr{\left\{e^{-\beta(\hat{H}-\mu%
\hat{N})} \BT_t \hat{A}_0(t) \hat{A}_0\right\}},  \label{dynamical_cf}
\end{equation}%
where $\mathcal{T}_{t}$ is the time ordering operator along real time and
\begin{equation}
\hat{A}_{0}(t)=e^{it(\hat{H}-\mu \hat{N})}\hat{A}_{0}e^{-it(\hat{H}-\mu \hat{%
N})}.\label{time-operator}
\end{equation}%
Here, $\hat{A}_0$ is the operator counterpart of $A_\bq(\tau)$ defined in Eq.~\eqref{a8} at $\bq=0$:
\begin{equation}
 \hat{A}_0 = \frac{1}{\sqrt{V}}\sum_\bp c_\bp^\dagger\check\Sigma_2c_\bp,
\end{equation}
where $c_\bp^\dagger$ and $c_\bp$ are fermion creation and destruction operators.

Since $\hat{A}_0$ conserves the number of particles,
this definition of time dependency is in fact equivalent to the Heisenberg representation of operators.
The correlation function $C^\BT(t)$ can be measured, for example, by magnetic neutron scattering which is sensitive to the magnetic fields produced by the fluctuating loop currents (see Appendix~\ref{section:scattering}).

Analogously, one can define imaginary time counterpart of Eq.~%
\eqref{dynamical_cf}:
\begin{equation}
C^{\mathrm{M}}(\tau )=\frac{1}{V\mathcal{Z}}\tr{\left\{e^{-\beta(\hat{H}-\mu%
\hat{N})} \BT_\tau \hat{A}_0(\tau) \hat{A}_0\right\}},  \label{imaginary_cf}
\end{equation}%
where $\mathcal{T}_{\tau }$ now denotes the ordering operator along the
imaginary time and
\begin{equation}
\hat{A}_{0}(\tau )=e^{\tau (\hat{H}-\mu \hat{N})}\hat{\mathcal{A}}%
_{0}e^{-\tau (\hat{H}-\mu \hat{N})}.
\end{equation}

By directly applying the time ordering operators, one can also re-express $%
C^{\mathcal{T}}(t)$ and $C^{\mathrm{M}}(\tau )$ as
\begin{equation}
C^{\mathcal{T}}(t)=\frac{\tr{\left\{e^{-\beta(\hat{H}-\mu\hat{N})}
e^{i|t|(\hat H - \mu\hat N)} \hat{A}_0 e^{-i|t|(\hat H-\mu\hat N)}
\hat{A}_0\right\}}}{V\mathcal{Z}}  \label{rt_cf}
\end{equation}%
\begin{equation}
C^{\mathrm{M}}(\tau )=\frac{\tr{\left\{e^{-\beta(\hat{H}-\mu\hat{N})}
e^{|\tau|(\hat H - \mu\hat N)} \hat{A}_0 e^{-|\tau|(\hat H-\mu\hat N)}
\hat{A}_0\right\}}}{V\mathcal{Z}}.  \label{it_cf}
\end{equation}

As one may notice from comparing Eq.~\eqref{rt_cf} and Eq.~\eqref{it_cf},
correlation functions $C^\mathcal{T}(t)$ and $C^\mathrm{M}(\tau)$ are
closely related: they can be obtained from each other by analytic
continuation~\cite{negele}. To describe it in more precise terms, let us
introduce the following two functions of the complex argument $z = \tau+it$:
\begin{equation}
C^>(z) = \frac{\tr{\left\{e^{-\beta(\hat H-\mu\hat N)} e^{z(\hat H - \mu\hat
N)}\hat{A}_0 e^{-z(\hat H -\mu\hat N)}\hat A_0\right\}}}{V\mathcal{Z}},
\end{equation}
\begin{equation}
C^<(z) = \frac{\tr{\left\{e^{-\beta(\hat H-\mu\hat N)} e^{-z(\hat H -
\mu\hat N)}\hat{A}_0 e^{z(\hat H -\mu\hat N)}\hat A_0\right\}}}{V\mathcal{Z}}%
.
\end{equation}
In the complex $z$ plane, the function $C^\mathrm{M}$ is defined on the real axis,
while the function $C^\mathcal{T}$ is defined on the imaginary axis. The
greater function $C^>(z)$ is the analytic continuation of both $C^\mathrm{M}%
(\tau)$ and $C^\mathcal{T}(t)$ into the first quadrant $\re{z}>0$, $\im{z}>0$%
. Conversely, the lesser function $C^<(z)$ is the analytic continuation of
both $C^\mathrm{M}(\tau)$ and $C^\mathcal{T}(t)$ into the third quadrant $%
\re{z}<0$, $\im{z}<0$:
\begin{align}
C^\mathrm{M}(\tau) = C^>(\tau) & \qquad \tau>0 \\
C^\mathrm{M}(\tau) = C^<(\tau) & \qquad \tau<0
\end{align}
\begin{align}
C^\mathcal{T}(t) = C^>(it) & \qquad t>0 \\
C^\mathcal{T}(t) = C^<(it) & \qquad t<0
\end{align}

In practice, the connection between real- and imaginary-time correlation
functions is established usually first by converting everything into the
frequency domain, and then performing the analytic continuation in the
transformed domain. We shall demonstrate, however, that the procedure of the
analytic continuation in the frequency domain is ambiguous for Instanton
Crystal State. As a result, the function $C^\mathcal{T}(t)$ should be
obtained from $C^\mathrm{M}(\tau)$ by analytic continuation directly in the
time domain.

In conclusion of this subsection, we would also like to point out that the
complex time plane is typically defined in such a way that the real time
corresponds to the real axis while imaginary time corresponds to the
imaginary axis of the complex plane. However, we found a complex time plane
definition with interchanged axes more suitable for the presented work.


\subsection{Analytic continuation in the frequency domain.\label{continuation-freq}}

To discuss contination in the frequency domain, we need to introduce yet another two real-time correlation functions, retarded and advanced one:
\begin{align}
 C^\mathrm{R}(t) & = -\frac{i\Theta(t)}{V\BZ}\ttr{e^{-\beta(\hat H - \mu\hat N)}\left[\hat{A}_0(t),\hat{A}_0\right]},\\
 C^\mathrm{A}(t) & = \frac{i\Theta(-t)}{V\BZ}\ttr{e^{-\beta(\hat H - \mu\hat N)}\left[\hat{A}_0(t),\hat{A}_0\right]},
\end{align}
where $\Theta(t)$ is the Heaviside theta-function.

For the real-time functions, we switch to the frequency domain by Fourier transform
\begin{equation}
 C^{\BT,\mathrm{R},\mathrm{A}}(\omega) = \lint_{-\infty}^{+\infty} dt e^{i\omega t-\eta|t|} C^{\BT,\mathrm{R},\mathrm{A}}(t).
\end{equation}
The transformed real-time functions are related to each other by \cite{agd}
\begin{equation}
\re{C^\BT(\omega)} = \mp\im{C^{\mathrm{R}/\mathrm{A}}(\omega)}\coth{\frac{\beta\omega}{2}},
\end{equation}
\begin{equation}
 \im{C^\BT(\omega)} = \re{C^{\mathrm{R}/\mathrm{A}}(\omega)}.
\end{equation}
On the other-hand, the imaginary-time function $C^\mathrm{M}(t)$ is expanded in Fourier series over Matsubara frequencies $\omega_n = 2\pi T n$, $n\in \mathbb{Z}$:
\begin{equation}
 C^\mathrm{M}(i\omega_n) = \lint_0^\beta d\tau e^{i\omega_n\tau} C^\mathrm{M}(\tau).
\end{equation}
Finally, the retarded correlation function $C^\mathrm{R}(\omega)$ is obtained from $C^\mathrm{M}(i\omega_n>0)$ by analytic continuation to the real axis in the upper complex frequency half plane: $C^\mathrm{R}(\omega) = -C^\mathrm{M}(\omega + i0)$. Analogously, the advanced correlation function $C^\mathrm{A}(\omega)$ is obtained from
$C^\mathrm{M}(i\omega_n<0)$ by analytic continuation to the real axis in the lower complex frequency half plane: $C^\mathrm{A}(\omega) = -C^\mathrm{M}(\omega - i0)$.

There are certain subtleties associated with continuation from a discrete set of Matsubara frequencies.
For concreteness, let us consider the analytic continuation in the upper complex half plane.
The accumulation point for the subset of Matsubara frequencies in the upper half plane is $+i\infty$.
Correspondingly, the uniqueness of the continuation procedure is normally guaranteed by the regular behaviour of $C^\mathrm{M}(i\omega_n)$ in the vicinity of the accumulation point, i.e.,
in the limit $i\omega_n\rightarrow +i\infty$.

This picture gets problematic if one considers the Instanton Crystal state.
If the Instanton Crystal consists of $m$ instanton-antiinstanton pairs, the period of the corresponding imaginary-time correlation function is $\beta/m$.
As a consequence, coefficients $C^\mathrm{M}(i\omega_n)$ are non-zero only for $n = mk$, $k\in\mathbb{Z}$.
Now, imagine we pick two subsequences of Matsubara frequencies, $i\omega_{n_k} = i\omega_{mk}$ and $i\omega^\prime_{n_k} = i\omega_{mk-1}$, and then perform the analytic continuation either using only the values $C^\mathrm{M}(i\omega_{n_k})$ or using only the values $C^\mathrm{M}(i\omega^\prime_{n_k})$.
On one hand, one expects from the Identity Theorem of Complex Analysis that the results of these two continuations should coincide with result of the analytic continuation with the full set of Matsubara frequencies $C^\mathrm{M}(i\omega_n)$. On the other hand, the result of continuation using $C^\mathrm{M}(i\omega^\prime_{n_k})$ is constant zero, while the result of continuation using $C^\mathrm{M}(i\omega_{n_k})$ is some non-trivial function.

This allows us to concude that analytic continuation in the frequency domain is ill defined in the case of Instanton Crystal state.

\subsection{Direct analytic continuation in time domain}

In the mean-field approximation, the imaginary-time correlation function $C^\mathrm{M}(\tau)$ has the form
\begin{multline}
C^{\mathrm{M}}(\tau )
= \\ \* =
\frac{4}{\beta }\int\limits_{0}^{\beta }d\tau _{0}%
\left[ \int \frac{d\mathbf{p}}{(2\pi )^{2}}\tr{\left\{\check\Sigma_2\check
G_\bp(\tau+\tau_0,\tau+\tau_0)\right\}}\right] \times \\
\*\times \left[ \int \frac{d\mathbf{p}^{\prime }}{(2\pi )^{2}}%
\tr{\left\{\check\Sigma_2\check G_{\bp^\prime}(\tau_0,\tau_0)\right\}}\right]
\label{corfunc_meanfield}
\end{multline}


Equation (\ref{corfunc_meanfield}) contains the averaging over the position $%
\tau _{0}$ of the instanton lattice.
This is the necessary procedure because the instanton crystal state configuration is invariant against the translations in the imaginary time $\tau$ (the periodic boundary conditions on the interval $\left(0,\beta\right)$ are implied), and one should properly average over the positions of the instanton lattice when calculating thermodynamic correlation functions.

In the following, it is convenient to introduce vector-function
\begin{equation}
\vec{S}_{\mathbf{p}}(\tau )=-\tr{\left\{\check{\vec\Sigma}\check
G_\bp(\tau,\tau)\right\}}  \label{vecc_momentum}
\end{equation}%
and
\begin{equation}
\vec{S}(\tau )=\int \frac{d\mathbf{p}}{(2\pi )^{2}}\vec{S}_{\mathbf{p}}(\tau
).  \label{vecc}
\end{equation}%
With the help of this definition we can rewrite Eq.~\eqref{corfunc_meanfield}
as
\begin{equation}
C^{\mathrm{M}}(\tau )=\frac{4}{\beta }\int\limits_{0}^{\beta }d\tau
_{0}S_{2}(\tau +\tau _{0})S_{2}(\tau _{0}).\label{corfunc_mf_continued}
\end{equation}%
The problem of analytic continuation of $C^{\mathrm{M}}(\tau )$ hence
reduces to the analytic continuation of the function $S_{2}(\tau +\tau _{0})$%
. Once it is done, the real-time function $C^{\mathcal{T}}(t)$ can be
expressed as
\begin{equation}
C^{\mathcal{T}}(t)=\frac{4}{\beta }\int\limits_{0}^{\beta }d\tau
_{0}S_{2}(\tau _{0}+it)S_{2}(\tau _{0}).
\end{equation}

The starting point for the continuation of $S_{2}(\tau )$ is the gap
equation~\eqref{gap:general}. The non-local kernel $K^{-1}(\tau -\tau
^{\prime }|\omega _{0})$ (see definition~\eqref{kernel_inverse}) is the
inverse to the direct kernel $K(\tau -\tau ^{\prime }|\omega _{0})$, which
is equal to (see \cite{se})
\begin{multline}
K(\tau -\tau ^{\prime }|\omega _{0})=(U_{0}+U_{2})\delta (\tau -\tau
^{\prime })- \\
\*-U_{2}K_{0}(\tau -\tau ^{\prime }|\omega _{0}).
\end{multline}%
Applying the direct kernel to both sides of Eq.~\eqref{gap:general}, we
rewrite it equivalently as
\begin{multline}
b(\tau )=(U_{0}+U_{2})\int \frac{d\mathbf{p}}{(2\pi )^{2}}S_{2\mathbf{p}%
}(\tau )- \\
\*-U_{2}\int\limits_{0}^{\beta }d\tau ^{\prime }K_{0}(\tau -\tau ^{\prime
}|\omega _{0})\int \frac{d\mathbf{p}}{(2\pi )^{2}}S_{2\mathbf{p}}(\tau
^{\prime }),  \label{gap:rewritten}
\end{multline}%
where we also used the definition~\eqref{vecc_momentum}.

Non-local kernel $K_{0}$ has an important property: it is the Green's
function for a certain differential operator with periodic boundary
conditions (see \cite{se}):
\begin{equation}
\left[ -\frac{d^{2}}{d\tau ^{2}}+\omega _{0}^{2}\right] K_{0}(\tau -\tau
^{\prime }|\omega _{0})=\omega _{0}^{2}\delta (\tau -\tau ^{\prime }).
\end{equation}%
Let us apply $[-\partial _{\tau }^{2}+\omega _{0}^{2}]$ to the both sides of
Eq.~\eqref{gap:rewritten}:
\begin{multline}
\frac{d^{2}b(\tau )}{d\tau ^{2}}+\omega _{0}^{2}b(\tau )=(U_{0}+U_{2})\int
\frac{d\mathbf{p}}{(2\pi )^{2}}\frac{d^{2}S_{2\mathbf{p}}(\tau )}{d\tau ^{2}}%
+ \\
\*+U_{0}\omega _{0}^{2}\int \frac{d\mathbf{p}}{(2\pi )^{2}}S_{2\mathbf{p}%
}(\tau ).  \label{gap:differential}
\end{multline}

The obtained differential equation needs to be supplemented by another one
which relates the derivatives of the vector-functions $\vec S_{\mathbf{p}%
}(\tau)$ to the order parameter field $b(\tau)$. To derive it, let us turn
our attention to the fermion Green's function $\check G_\mathbf{p}%
(\tau,\tau^\prime)$. In addition to Eq.~\eqref{greens_function}, it also
satisfies
\begin{multline}
\check G_\mathbf{p}(\tau,\tau^\prime)\left[(-\overset{\leftarrow}{\partial}%
_{\tau^\prime} + \varepsilon_\mathbf{p}^+)\check{\mathrm{I}} + \varepsilon_%
\mathbf{p}^-\check\Sigma_3 - b(\tau)\check\Sigma_2\right] = \\
\* = -\check{\mathrm{I}}\delta(\tau-\tau^\prime)
\label{greens_function_adjoint}
\end{multline}
Subtracting Eq.~\eqref{greens_function_adjoint} from Eq.~%
\eqref{greens_function} and putting $\tau^\prime = \tau$, one obtains
\begin{multline}
\left.\left(\partial_\tau + \partial_{\tau^\prime}\right)\check G_\mathbf{p}%
(\tau,\tau^\prime)\right|_{\tau^\prime=\tau} \equiv \partial_\tau \check G_%
\mathbf{p}(\tau,\tau) = \\
\* = -[\varepsilon_\mathbf{p}^-\check\Sigma_3 - b(\tau)\check\Sigma_2,
\check G_\mathbf{p}(\tau,\tau)].  \label{greens_function_commutator}
\end{multline}
Now, let us multiply both sides of equation~%
\eqref{greens_function_commutator} by $\check\Sigma_i$, $i=1,2,3$, and apply
trace. Using the cyclic property of trace and commutation relations for
Pauli matrices, one can write the resulting equations jointly as
\begin{equation}
\frac{d \vec{S}_{\mathbf{p}}(\tau)}{d\tau} = -2i \vec B_\mathbf{p}%
(\tau)\times \vec{S}_\mathbf{p}(\tau),  \label{pseudospins}
\end{equation}
where
\begin{equation}
\vec B_\mathbf{p}(\tau) = \left(0,\quad -b(\tau),\quad \varepsilon_\mathbf{p}%
^-\right).  \label{pseudospin_field}
\end{equation}

The necessary final touch is to get rid of the second derivative of $C_{2%
\mathbf{p}}(\tau)$ in Eq.~\eqref{gap:differential}. This can be achieved by
applying Eq.~\eqref{pseudospins} twice and leads to
\begin{multline}
\frac{d^2 b(\tau)}{d\tau^2} = \omega_0^2\left[U_0 \int\frac{d\mathbf{p}}{%
(2\pi)^2}S_{2\mathbf{p}}(\tau) - b(\tau)\right] + \\
\* + (U_0+U_2)\int \frac{d\mathbf{p}}{(2\pi)^2} 4\varepsilon_\mathbf{p}^-%
\left[b(\tau)S_{3\mathbf{p}}(\tau) + \varepsilon_\mathbf{p}^-S_{2\mathbf{p}%
}(\tau)\right].  \label{gap:diff_final}
\end{multline}

Equations~\eqref{pseudospins} and~\eqref{gap:diff_final} constitute a closed
system which the functions $b(\tau )$, $db(\tau )/d\tau $ and $\vec{S}_{%
\mathbf{p}}(\tau )$ satisfy: once the values of $b(\tau )$, $db(\tau )/d\tau
$ and $\vec{S}_{\mathbf{p}}(\tau )$ at some $\tau $ are known, the functions
at other points can be reconstructed by integrating Eqs.~\eqref{pseudospins}
and~\eqref{gap:diff_final}. Correspondingly, the analytic continuation of
these functions can be facilitated by analytic continuation of the system of
Eqs.~\eqref{pseudospins} and~\eqref{gap:diff_final}. 


After the analytic continuation of equations~\eqref{pseudospins} and~%
\eqref{gap:diff_final} are done, it is convenient to restrict them to some
line parallel to the imaginary axis (with our definition of complex time,
real time runs along the imaginary axis). The equations along this line,
which can be parametrized as $\tau +it$ with fixed $\tau $, are obtained
from equations~\eqref{pseudospins} and~\eqref{gap:diff_final} by replacing
\begin{equation}
\frac{df(\tau )}{d\tau }\rightarrow -i\frac{df(\tau +it)}{dt},
\end{equation}%
where $f(\tau )$ is one of the functions $b(\tau )$ and $\vec{S}_{\mathbf{p}%
}(\tau )$.

The resulting equations take the form
\begin{equation}
\frac{d\vec{S}_{\mathbf{p}}(\tau +it)}{dt}=2\vec{B}_{\mathbf{p}}(\tau
+it)\times \vec{S}_{\mathbf{p}}(\tau +it),  \label{pseudospins_rt}
\end{equation}%
\begin{multline}
\frac{d^{2}b(\tau +it)}{dt^{2}}=\omega _{0}^{2}\left[ b(\tau +it)-U_{0}\int
\frac{d\mathbf{p}}{(2\pi )^{2}}S_{2\mathbf{p}}(\tau +it)\right] - \\
\*-(U_{0}+U_{2})\int \frac{d\mathbf{p}}{(2\pi )^{2}}4\varepsilon _{\mathbf{p}%
}^{-}\left[ b(\tau +it)S_{3\mathbf{p}}(\tau +it)+\right. \\
\*+\left. \varepsilon _{\mathbf{p}}^{-}S_{2\mathbf{p}}(\tau +it)\right] .
\label{gap:diff_final_rt}
\end{multline}%
These equations allow one to determine $b(\tau +it)$, $db(\tau +it)/dt$ and $%
\vec{S}_{\mathbf{p}}(\tau +it)$ from their values at $t=0$. Moreover, the
initial value of $db(\tau +it)/dt$ at $t=0$ can be related to the values of $%
b(\tau )$ and $\vec{S}_{\mathbf{p}}(\tau )$. To do that, let us
differentiate Eq.~\eqref{gap:general} once:
\begin{multline}
\frac{db(\tau )}{d\tau }+\frac{U_{2}}{U_{0}}\int\limits_{0}^{\beta }d\tau
^{\prime }\frac{dK_{0}(\tau -\tau ^{\prime }|\tilde{\omega}_{0})}{d\tau }%
b(\tau ^{\prime })= \\
=(U_{0}+U_{2})\int \frac{d\mathbf{p}}{(2\pi )^{2}}\frac{dS_{2\mathbf{p}%
}(\tau )}{d\tau }= \\
=-2i(U_{0}+U_{2})\int \frac{d\mathbf{p}}{(2\pi )^{2}}\varepsilon _{\mathbf{p}%
}^{-}S_{1\mathbf{p}}(\tau ).
\end{multline}%
Here, we used Eq.~\eqref{pseudospins} in the last line. Then, we can write
\begin{multline}
\left. \frac{db(\tau +it)}{dt}\right\vert _{t=0}=i\frac{db(\tau )}{d\tau }=
\\
=i\frac{U_{2}}{U_{0}}\int\limits_{0}^{\beta }d\tau ^{\prime }\frac{%
dK_{0}(\tau -\tau ^{\prime }|\tilde{\omega}_{0})}{d\tau }b(\tau ^{\prime })+
\\
+2(U_{0}+U_{2})\int\frac{d\mathbf{p}}{(2\pi )^{2}}%
\varepsilon _{\mathbf{p}}^{-}S_{1\mathbf{p}}(\tau ).\label{derivative_init}
\end{multline}%
%
%
%
%
%
%
%
%
%
%
%

\section{Analytical results for the real-time order parameter correlation function in the mean-field approximation at $U_2=0$.\label{section:analytics}}



\begin{figure*}[t]
 \includegraphics[width=5.7in]{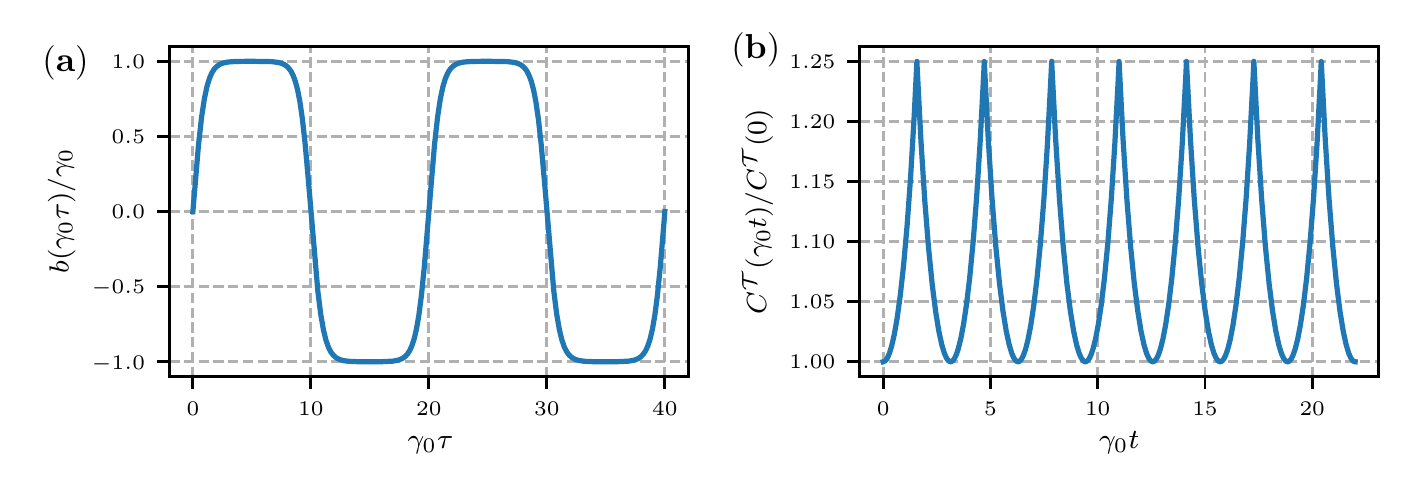}
 \caption{
 Instanton lattice configuration and the corresponding real-time correlation function at $U_2=0$ and zero temperature.
 (a) Normalized order parameter of the instanton lattice $b(\gamma_0 \tau)/\gamma_0$ as the function of dimensionless imaginary time $\gamma_0 \tau$. The dimensionless period of instanton lattice is $\gamma_0 W = 20.0$.
 (b) Normalized correlation function of the order parameter $C^\BT(\gamma_0 t)/C^\BT(0)$ as the function of dimensionless real time $\gamma_0 t$ at $U_2 = 0$.}
 \label{jacobi_continuation}
\end{figure*}

In the previous section, we introduced the correlation functions of the order parameter both on the real-time axis, $C^\mathcal{T}(t)$, Eq.~\eqref{dynamical_cf}, and on the imaginary-time axis, $C^\mathrm{M}(\tau)$, Eq.~\eqref{imaginary_cf}.
Now, we are going to calculate these functions at $U_2=0$.
Of course, we have shown previously~\cite{se} that the Instanton Crystal state is not thermodynamically stable in the absence of non-local interaction.
However, we have also seen~\cite{se} that even a small non-local interaction $U_2$ stabilizes the Instanton Crystal.
As a result, calculations presented in this section allow us to demonstrate explicitly how the analytic continuation from $\tau$ to $t$ is performed and how the instanton-antiinstanton train in imaginary time is transformed into oscillations in real time. In addition to that, for the physically relevant case of $U_2/U_0\ll 1$, we may expect that the results are close to the results at $U_2=0$.

With the help of the mean-field equation~\eqref{gap:general} in the absence of non-local interaction,
the correlation function $C^{\mathrm{M}}(\tau )$, Eq.~\eqref{corfunc_mf_continued}, can be
written in the form of a correlation function of the order parameter $b\left( \tau
\right)$:
\begin{equation}
U_0^2C^{\mathrm{M}}(\tau )=\frac{1}{\beta}\int_{0}^{\beta }b\left(\tau +\tau
_{0}\right) b\left( \tau _{0}\right) d\tau _{0}.  \label{p1}
\end{equation}%
If we take into account the periodic properties of $b(\tau)$, this leads to the expression
\begin{equation}
U_0^2C^{\mathrm{M}}(\tau )=\frac{1}{W}\int_{0}^{W}b\left( \tau +\tau
_{0}\right) b\left( \tau _{0}\right) d\tau _{0},  \label{p2}
\end{equation}%
where the period $W$ of the instanton-antiinstanton chain is given by Eq.~\eqref{p0}.
The analytical continuation $\tau \rightarrow \tau +it$ in Eq.~\eqref{p2} is
straightforward and we write
\begin{equation}
U_0^2 C^{\mathrm{M}}(\tau +it)=\frac{1}{W}\int_{0}^{W}b\left(\tau + it +\tau
_{0}\right) b\left( \tau _{0}\right) d\tau _{0}.  \label{p3}
\end{equation}%

Now, we use a known Fourier series expansion for Jacobi elliptic function $\sn{(x|k)}$~\cite{as} to write the function $b\left( \tau \right)$, Eq.~\eqref{jacobi_order_parameter}, as
\begin{equation}
b\left( \tau \right) =\frac{4\pi }{W}\sum_{l=0}^{\infty }\frac{\sin \frac{%
2\pi \left( 2l+1\right) \tau }{W}}{\sinh \frac{\pi \left( 2l+1\right)
W^{\prime }}{W}},  \label{p4}
\end{equation}
where%
\begin{equation}
W^{\prime }=2K\left( k^{\prime }\right) /\gamma,\qquad {k^\prime}^2 = 1- k^2 .\label{p5}
\end{equation}%

The series expansion~\eqref{p4} can be continued to the strip $\tau+it$, $-W^\prime/2<t<W^\prime/2$.
This allows us to rewrite Eq.~\eqref{p3} in the form%
\begin{multline}
 U_0^2 C^{\mathrm{M}}(\tau +it)
 = \\ =
 \frac{16\pi ^{2}}{W^{3}}\sum_{l,l^{\prime }=0}^{\infty }\frac{1}{\sinh
\frac{\pi \left( 2l+1\right) W^{\prime }}{W}\sinh \frac{\pi \left(
2l^{\prime }+1\right) W^{\prime }}{W}}
\times \\ \times
\int_{-W/2}^{W/2}\sin \frac{2\pi \left( 2l+1\right) \left( \tau
_{0}+\tau +it\right) }{W}\sin \frac{2\pi \left( 2l^{\prime }+1\right) \tau
_{0}}{W}d\tau _{0}. \label{p6}
\end{multline}
%
Calculating the integral over $\tau _{0}$ for $l,l^{\prime }\geq 0$, we
obtain
\begin{equation}
 C^{\mathrm{M}}(\tau +it) = \frac{8\pi ^{2}}{W^{2}}\sum_{l=0}^{\infty }\frac{\cos \frac{2\pi \left(
2l+1\right) \left( \tau +it\right) }{W}}{\sinh ^{2}\frac{\pi \left(
2l+1\right) W^{\prime }}{W}}. \label{p7}
\end{equation}
%
It is clear that the function $C^{\mathrm{M}}(\tau +it)$ is analytic in the
vicinity of zero, and the method of the analytical continuation from the
imaginary time $\tau $ to the real one $t$ works. At the same time, although
the function $C^{\mathrm{M}}(\tau +it)$, Eq.~\eqref{p7}, is explicitly
periodic in $\tau $, its periodicity is not evident in real time $t$.

The periodicity and divergencies of the Jacobi elliptic functions are well-known.
As such, $b(\tau + it)$ has the poles at $iW^\prime/2$ and $W/2 + iW^\prime/2$, and the period along $t$ equals $W^\prime$.
Hence, we can represent the function $C^{\mathcal{T}}(t)$ in the form of the
Fourier series%
\begin{equation}
C^{\mathcal{T}}(t)=\sum_{l=0}^{\infty }\sum_{L=0}^{\infty }C_{L,l}\cos
\left( \frac{2\pi Lt}{W^{\prime }}\right) .  \label{p8}
\end{equation}%
Setting $\tau =0$ in Eq.~\eqref{p7}, we calculate the coefficients $C_{L,l}$ to be
\begin{equation}
C_{L,l}=\frac{8\pi W^{\prime }}{W^{3}}\frac{\left( -1\right) ^{L}}{\sinh
\frac{\pi \left( 2l+1\right) W^{\prime }}{W}}\frac{2l+1}{\left( 2l+1\right)
^{2}+L^{2}},  \label{p9}
\end{equation}%
and obtain the final expression for the correlation function%
\begin{multline}
C^{\mathcal{T}}(t) =\frac{8\pi W^{\prime }}{W^{3}}\sum_{l=0}^{\infty
}\sum_{L=0}^{\infty }\left( -1\right) ^{L}\frac{1}{\sinh \frac{\pi \left(
2l+1\right) W^{\prime }}{W}} \times  \\
\times \frac{2l+1}{\left( 2l+1\right) ^{2}+L^{2}}\cos \left( \frac{2\pi Lt%
}{W^{\prime }}\right) .\label{p10}
\end{multline}%
In the limit $k\rightarrow 1$ of the large period of the
instanton-antiinstanton lattice, $K\left( k^{\prime }\right)$ $\rightarrow
\pi /2$, and Eq.~\eqref{p5} transforms into
\begin{equation}
W^{\prime }\rightarrow \pi /\gamma .  \label{p11}
\end{equation}%
We see from Eqs.~\eqref{p10} and~\eqref{p11} that, in this limit, the correlation
function $C^{\mathcal{T}}(t)$ is periodic in real time with the frequency $%
2\gamma .$ This frequency equals to the excitation gap in SFMOHS studied
previously   \cite{volkov1, volkov2, volkov3}.

To illustrate this discussion, we provide Fig.~\ref{jacobi_continuation},
where we present an example of instanton lattice configuration at $U_2=0$
and the corresponding real-time correlation function obtained after analytic continuation.
Panel (a) displays the normalized order parameter $b(\gamma_0 \tau)/\gamma_0$ as the function of dimensionless imaginary time $\gamma_0 \tau$. Here, $\gamma_0$ stands for the static gap at zero temperature (see Eq.~\eqref{self_consistency:static}).
The order parameter configuration shown corresponds to the Instanton Crystal with dimensionless period $\gamma_0 W = 20.0$.
Panel (b) displays the normalized order parameter correlation function $C^\BT(\gamma_0 t)/C^\BT(0)$ as the function of the dimensionless real time $\gamma_0 t$.

\section{Numerical Results\label{section:numerics}}

\begin{figure*}[t]
 \includegraphics[width=7in]{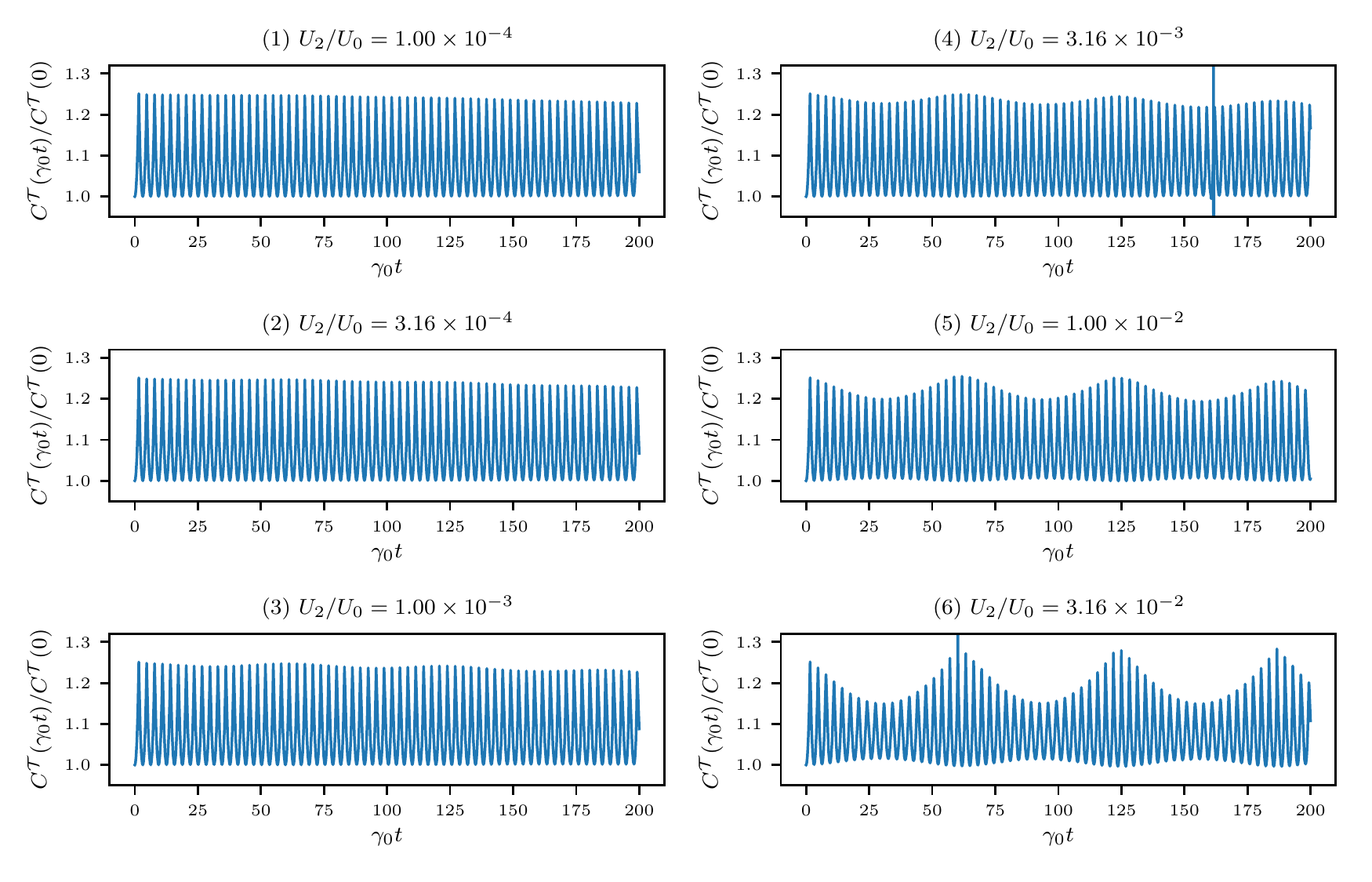}
 \caption{Normalized real-time correlation function of the order parameter $C^\BT(\gamma_0 t)/C^\BT(0)$ for fixed $\omega_0/gamma_0 = 0.10$ and $\gamma_0 W=20.0$ but for varying strength of non-local interaction $U_2/U_0$.}
 \label{u2runs}
\end{figure*}

\begin{figure*}[t]
 \includegraphics[width=7in]{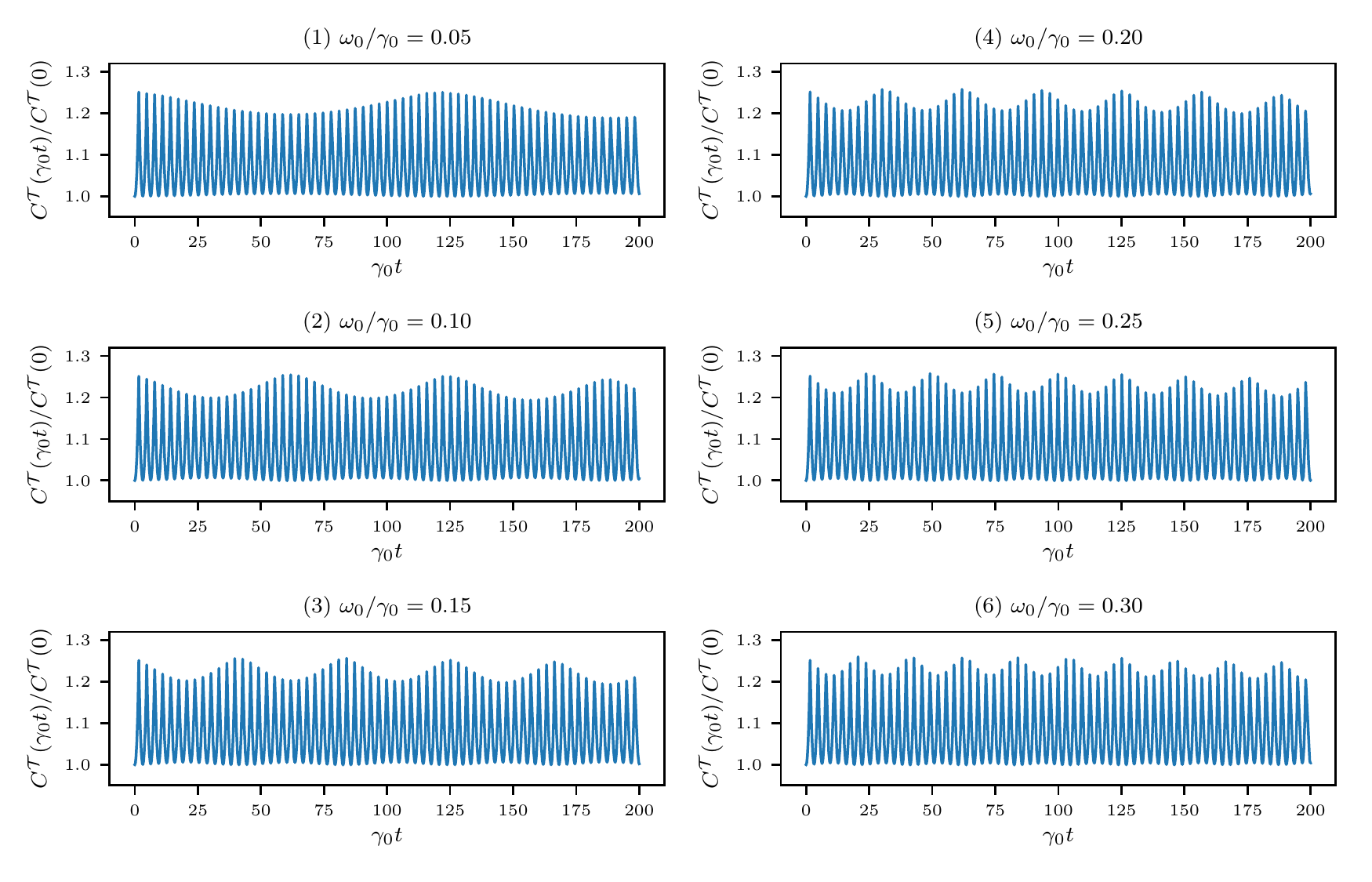}
 \caption{Normalized real-time correlation function of the order parameter $C^\BT(\gamma_0 t)/C^\BT(0)$ for fixed $U_2/U_0 = 0.10$ and $\gamma_0 W=20.0$ but for varying frequency of bosonic mode $\omega_0/\gamma_0$.}
 \label{omegaruns}
\end{figure*}

\begin{figure*}[t]
 \includegraphics[width=7in]{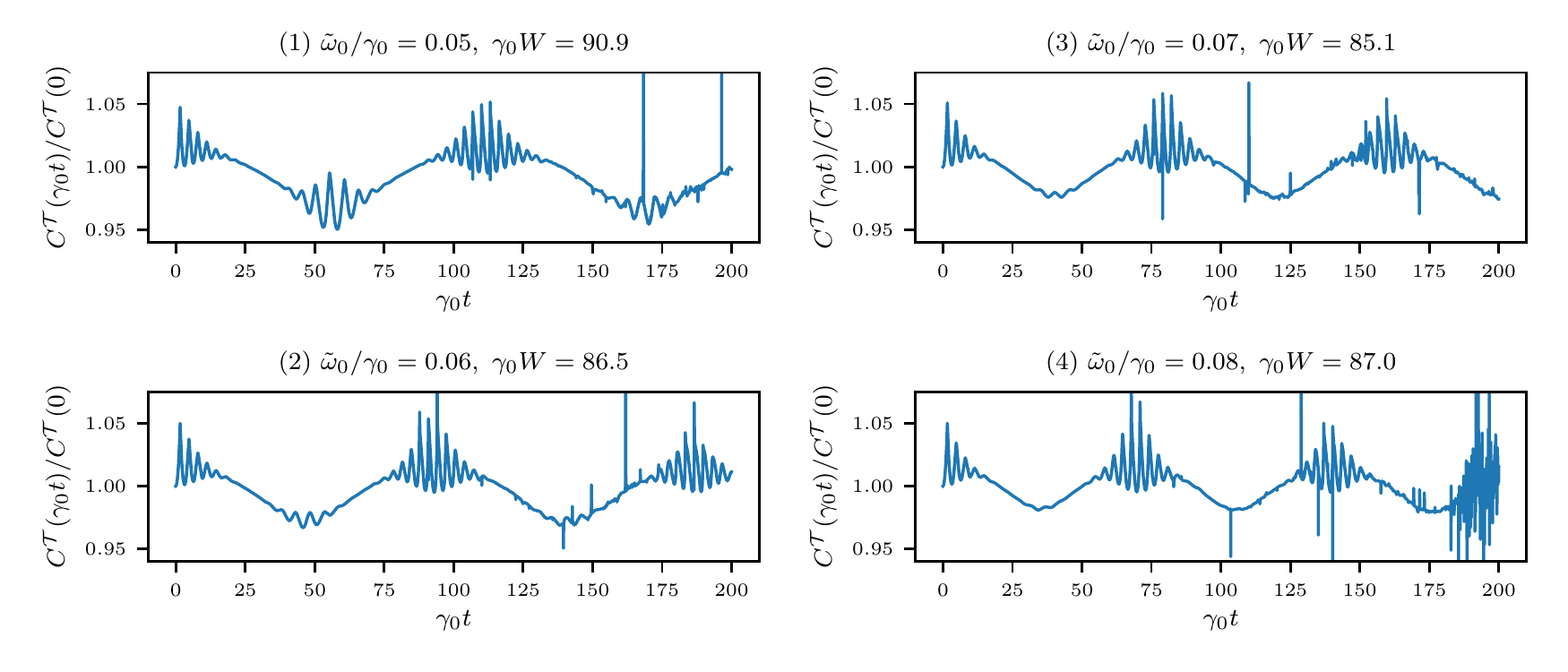}
 \caption{Normalized real-time correlation function of the order parameter $C^\BT(\gamma_0 t)/C^\BT(0)$ for fixed $U_2/U_0 = 0.20$ and varying modified frequency of the bosonic mode $\tilde\omega_0/\gamma_0$. The period of the instanton lattice was determined from the minimum of the free energy.}
 \label{physicalruns}
\end{figure*}

In order to proceed with the actual numerical computations, we had to specify particular fermion dispersions $\varepsilon_{1,2}(\bp)$ (see Eq.~\eqref{a4}).
In agreement with our previous work~\cite{se}, we chose it in the same form as it appears in SFMOHS:
\begin{equation}
 \varepsilon_1(\bp) = \alpha p_x^2 - \beta p_y^2 - \mu,
 \quad \varepsilon_2(\bp) = \alpha p_y^2 - \beta p_x^2 -\mu,
\end{equation}
where $\mu$ is the chemical potential.
We also introduced an energy cutoff $\Lambda$ limiting the width of the dispersion:
\begin{equation}
 \frac{\alpha+\beta}{2}(p_x^2 + p_y^2)<\Lambda.
\end{equation}
The specific values used throughout the computations were $\alpha = \beta = 1.0$, $\Lambda/\gamma_0 = 1.0$, $\mu/\gamma_0 = 0.0$.


The configuration of the Instanton Crystal can be controlled by the following set of dimensionless parameters: $(U_0+U_2)/\gamma_0$, the ratio $U_2/U_0$, the period of the lattice $\gamma_0 W$, the modified frequency of the boson mode $\gamma_0\tilde\omega_0$ and the number of the periods $m$.
Here, $\gamma_0$ stands for the solution of the static gap equation~\eqref{self_consistency:static} at zero temperature and in the absence of non-local term ($U_2=0$).
We also used the static self-consistency equation~\eqref{self_consistency:static} at zero temperature to fix the value of the parameter $(U_0+U_2)/\gamma_0$.

These parameters are not completely independent.
The period of the lattice $\gamma_0 W$ and their number $m$ need to be determined from the dimensionless inverse temperature $\gamma_0\beta$ and other parameters by the condition of the minimum of the free-energy functional~\eqref{fe_functional}. Correspondingly, in the limit of zero temperature, $m\rightarrow+\infty$ and we determine only $\gamma_0 W$ from the same condition.

Despite these facts, it is convenient to treat these parameters as if they are independent for the purposes of the numeric investigation:
Generally, we expect the properties of the Instanton Crystal configuration to vary smoothly with the control parameters, while the phase transitions correspond to the jumps in the parameter space.

Finally, we should mention that throughout the numerical calculations we assumed the limit of zero temperature.

In Fig.~\ref{u2runs}, we display the computed normalized real-time correlation functions $C^\BT(\gamma_0 t)/C^\BT(0)$ for varying values of the strength of the non-local interaction $U_2/U_0$. The values of the other parameters were kept fixed: $\gamma_0 W = 20.0$, $\omega_0/\gamma_0=0.10$. The values of $U_2/U_0$ used for the calculations are listed in the figure.
As we may observe, the oscillations with period $W^\prime$ (see Eq.~\eqref{p5}) which were present in the limit $U_2/U_0=0$ survive for small but finite values of $U_2/U_0$.
At the same time, the amplitude of these surviving oscillations gets modulated, and the magnitude of the modulation grows together with the strength of the non-local interaction $U_2/U_0$.

It is natural to assume that the frequency of the modulation corresponds to the bosonic mode frequency. In Fig.~\ref{u2runs}, the period of the modulation does not change with variation of $U_2/U_0$ which correlates with the fact that we kept $\omega_0/\gamma_0$ fixed. To test this hypothesis, we performed another series of calculations where the strength of the non-local interaction $U_2/U_0$ was fixed while we changed the frequency of the bosonic mode $\omega_0/\gamma_0$.
The results are displayed in Fig.~\eqref{omegaruns}.
The period of the instanton lattice was again fixed at $\gamma_0 W=20.0$ while we kept $U_2/U_0 = 0.1$.
The values of $\omega_0/\gamma_0$ are listed in the figure.
As we may observe, the frequency of the modulation does indeed grow together with the bosonic frequency $\omega_0/\gamma_0$.

Finally, we performed a series of more physically realistic computations where the instanton lattice period was not chosen arbitrarily but was determined from the condition of the minimum of the free energy. The results are presented in Fig.~\ref{physicalruns}.
We kept the strength of the non-local interaction fixed at $U_2/U_0=0.2$ while the bosonic frequency was varied. The values of the modified bosonic frequency $\tilde\omega_0/\gamma_0$ and of the instanton lattice period $\gamma_0 W$ are listed in the figure as usual. In principle, we may again observe the structure of modulated oscillations with period $W^\prime$ which we noted in Figs.~\ref{u2runs} and~\ref{omegaruns}.
At the same time, it is important to point out that the modulation dominates the oscillations in this case.

\section{Validity of the \textquotedblleft No-go\textquotedblright\
theorem in the case of Instanton Crystal.\label{section:validity_no_go}}

In the publication \cite{watanabe} by Watanabe and Oshikawa (WO) a theorem
has been proposed that the time crystal is not possible in the thermodynamic
equilibrium and therefore one can speak of this phenomenon only including
non-equilibrium processes like, e.g., pumping, many body localization, etc.
(for a recent review, see \cite{sondhi3}).
There are many definitions of the
time crystals but we follow just the one used in Ref. \cite{watanabe}. The
arguments of \cite{watanabe} are based on scaling and counting powers of the
volume $V$ entering physical quantities.
Moreover, the proof of Watanabe and Oshikawa (WO) excludes even the possiblitiy of the time-dependence of the order parameter correlation function.
However, there are certain problems in their proof which render the theorem not applicable to the case of the instanton crystal.


WO consider a correlation function of two quite general operators $\hat{A}$
and $\hat{B}$ composed of local operators $a\left( \mathbf{r}\right) $ and $%
b\left( \mathbf{r}\right) $ in a form of integrals over the space
\begin{equation}
\hat{A}=\frac{1}{V}\int_{V}a\left( \mathbf{r}\right) d^{d}\mathbf{r,\quad }%
\hat{B}=\frac{1}{V}\int_{V}b\left( \mathbf{r}\right) d^{d}\mathbf{r,}
\label{ng1}
\end{equation}%
(we have normalized here the operators $\hat{A}$ and $\hat{B}$ multiplying
the integrals over $\mathbf{r}$ by $V^{-1}$). With this choice, the
operators $\hat{A}$ and $\hat{B}$ scale with the volume as $V^{0}$ and,
using the locality of the operators $a\left( \mathbf{r}\right) $ and $%
b\left( \mathbf{r}\right) $, one can assume that the commutators containing
the Hamiltonian $\hat{H}$ scale as
\begin{equation}
\left[ \hat{A},\hat{H}\right] \propto V^{0},\quad \left[ \hat{B},\hat{H}%
\right] \propto V^{0},\text{ }\left[ \left[ \hat{A},\hat{H}\right] ,\hat{A}%
\right] \propto V^{-1},  \label{ng3}
\end{equation}%
where $\left[ ,\right] $ are commutators. According to the assumption by WO,
the Hamiltonian $\hat{H}$ describes a completely general model with finite
range interactions, and one can make estimates simply starting with Eqs. (%
\ref{ng1}, \ref{ng3}) without assuming anything else.

\subsection{Zero temperature.}

The main part of the proof of the theorem by WO is based on the use of the
Cauchy-Schwarz inequality applied at zero temperature $T=0$ to a general
model with a general Hamiltonian $\hat{H}.$ The main claim of the theorem is
that non-decaying real-time dependence is impossible in any macroscopic
thermodynamically stable system. According to the authors, the restriction $%
T=0$ should not be so important, and they suggest in the second part of the
paper a discussion at finite temperature based on the rigorous bound for
spin systems~\cite{lieb}.

However, making scaling with $V$ and putting $T=0$ is not an unambiguous
procedure and the result depends on the problem considered. Generally, one
can take either the limit 1) $T\rightarrow 0$ and then $V\rightarrow \infty $
or 2) first $V\rightarrow \infty $ and then $T\rightarrow 0.$ What they have
in mind is apparently the case 1) when temperature $T$ can be smaller that
the energy of the first higher level.

If so, one should assume that in the limit $V\rightarrow \infty $ the first
excited state is separated from the ground state by a finite gap $\Delta
_{0} $ that does not vanish in the limit $V\rightarrow \infty .$ Certainly,
geometrically finite systems having a finite distance between the levels
exist and the limit $T\ll \Delta $ is possible but one cannot scale time
with volume in the way as it was done by WO. In any case, the model of
fermions and bosonic modes considered by us cannot be described by first
putting $T=0$ and only then $V\rightarrow \infty $. We start with a model of
a metal with a level separation $\Delta =\left( \nu V\right) ^{-1},$ where $%
\nu $ is the density of states, and fix a finite $T,$ which means that we
consider explicitly the case 2). Studying properties of the system in the
thermodynamic limit $V\rightarrow \infty $ we cannot keep the restriction $%
T\ll \Delta $ because $\Delta \rightarrow 0.$ Therefore the proof at $T=0$
presented in Ref. \cite{watanabe} cannot be applied in our model, and the
Cauchy-Schwarz inequality does not help.

We already considered the real-time correlation functions $%
C^{>}\left( z\right) $ and $C^{<}\left( z\right)$ in section~\ref{section:cont-introduction}. For comparison with
experiments, it is convenient to use their combination $C^\mathrm{K}\left( t\right) $:
\begin{multline}
 C^\mathrm{K}(t) = C^{>}(t) + C^{<}(t)
 = \\  =
 \frac{1}{\BZ}\ttr{e^{-\beta\hat{H}_g}\{\hat{A}(t),\hat{A}(0)\}},
\end{multline}
where $\hat{H}_g = \hat{H}-\mu \hat{N}$ and $\hat{A}(t)$ is given by Eq.~\eqref{a1}.
As we calculate thermodynamic averages, we also need operators
depending on the imaginary time $\tau $:
\begin{equation}
\hat{A}\left( it\right) =e^{i\hat{H}_gt}\hat{A}e^{-i\hat{H}_gt}.  \label{ng8}
\end{equation}

Estimating bounds is not what is usually done in physics. In this section we
simply want to understand in simple terms whether our results on the
dependence of the function $C^\mathrm{K}\left( t\right) $ on time $t$ really contradict
to the \textquotedblleft no-go\textquotedblright\ theorem \cite{watanabe}
that forbids any time dependence of the function $C^\mathrm{K}\left( t\right) $ in the
limit $V\rightarrow \infty $ or it is a special property of the instanton
crystal.

In order to answer this question, we simply consider the limit of short
times $t$ and expand the function $C^\mathrm{K}\left( t\right) $ in $t$.

The first terms of the expansion equal
\begin{multline}
C^\mathrm{K}\left( t\right) -C^\mathrm{K}\left( 0\right) =\frac{1}{\BZ}\sum_{n}e^{-\beta E_{n}}  \times \\
\times \left[ -\frac{it}{2}\left\langle n\left\vert \left[ \hat{H}_g,\hat{A}%
\right] \hat{A}\right\vert n\right\rangle -\frac{it}{2}\left\langle
n\left\vert \hat{A}\left[ \hat{H}_g,\hat{A}\right] \right\vert n\right\rangle %
\right]  - \\
-\frac{t^{2}}{2}\left\langle n\left\vert \left[ \hat{H}_g,\hat{A}\right]
^{2}\right\vert n\right\rangle .  \label{ng26}
\end{multline}%
The linear in $t$ terms cancel each other as they should. The quadratic in $%
t $ term is more interesting. Using the standard Heisenberg equation in the
imaginary time%
\begin{equation}
\frac{d\hat{A}\left( \tau \right) }{d\tau }=\left[ \hat{H}_g,\hat{A}\left(
\tau \right) \right] ,  \label{ng20}
\end{equation}%
we bring Eq.~(\ref{ng26}) to the form
\begin{multline}
C^\mathrm{K}\left( t\right) -C^\mathrm{K}\left( 0\right) \approx -\frac{t^{2}}{2\BZ}%
\sum_{n}e^{-\beta E_n}\left\langle n\left\vert \left[
\hat{H}_g,\hat{A}\right] ^{2}\right\vert n\right\rangle =  \\
=-\frac{t^{2}}{2\BZ}\sum_{n}e^{-\beta E_n}\left\langle n\left\vert \left[ \hat{H}_g,\hat{A}\left( \tau \right) \right]
^{2}\right\vert n\right\rangle  = \\
=-\frac{t^{2}}{2\BZ}\sum_{n}e^{-\beta E_n}\left\langle n\left\vert \frac{d\hat{A}\left( \tau \right) }{d\tau }\frac{d%
\hat{A}\left( \tau \right) }{d\tau }\right\vert n\right\rangle.  \label{ng27}
\end{multline}%
Now, using Eq.~\eqref{ng1} and introducing a correlation function $L\left(
\mathbf{r-r}^{\prime }\right) $
\begin{multline}
L\left( \mathbf{r-r}^{\prime }\right)  = \\
=\frac{1}{\BZ}\mathrm{Tr}\left[ e^{-\beta \hat{H}_g}\frac{da\left(
\tau ,\mathbf{r}\right) }{d\tau }\frac{da\left( \tau ,\mathbf{r}^{\prime
}\right) }{d\tau }\right]  \label{ng16}
\end{multline}%
we bring Eq.~(\ref{ng27}) to the form
\begin{equation}
C^\mathrm{K}\left( t\right) -C^\mathrm{K}\left( 0\right) =-\frac{t^{2}}{2}\frac{1}{V}\int L\left(
\mathbf{r-r}^{\prime }\right) d\mathbf{r}^{\prime }.  \label{ng28}
\end{equation}

The function $a\left( \tau ,\mathbf{r}\right) $ in our model corresponds to
the local loop currents which are local quantities. Local quantities are
also assumed in Ref. \cite{watanabe}. At first glance, this should lead to a
finite value of the integral over $\mathbf{r}^{\prime }$ in Eq. (\ref{ng28})
in any thermodynamic system, which would give in the limit $V\rightarrow
\infty $ the zero value of $C^\mathrm{K}\left( t\right) -C^\mathrm{K}\left( 0\right)$.

The only possibility to avoid this scenario is formation of a long range
order in space such that the correlation function $L\left( \mathbf{r}\right)
\rightarrow const$ in the limit $\left\vert \mathbf{r}\right\vert
\rightarrow \infty .$ Then, the integration over $\mathbf{r}^{\prime }$ in
Eq. (\ref{ng28}) would give an additional volume $V$ and the $C^\mathrm{K}\left(
t\right) -C^\mathrm{K}\left( 0\right) $ would become time-dependent. At the same time,
the function $L\left( \mathbf{r-r}^{\prime }\right) $, Eq. (\ref{ng16}),
contains finite derivative of the order parameter $a\left( \tau \right) $
with respect to the imaginary time $\tau$. To the best of our knowledge,
only the instanton crystal proposed in Ref. \cite{se} possesses this
property.

Of course, the consideration suggested in this section is not sufficient for
proving the existence of the slowly decaying real time oscillations but it does
indicate what has been missed in the arguments presented in Ref. \cite%
{watanabe}. For the explicit proof, one should perform calculations like
those carried out in the previous sections.

We emphasize that the very possibility of the existence of the slowly decaying
real time oscillations is the consequence of the existence of the instanton
crystal and one cannot expect them in other systems.

\subsection{Finite temperature.}

The proof of second part of \textquotedblleft no-go\textquotedblright\
theorem of Ref. \cite{watanabe}, which deals with the case of finite temperature, is based on the exact Lieb-Robinson bound for spin systems~\cite{lieb}.
Crucially, Watanabe and Oshikawa assume that Lieb-Robinson bound necessarily implies the existence of a similar bound in the Fourier representation.
However, as it is pointed out in~\cite{sondhi3}, this assumption is incorrect because Lieb-Robinson bound is satisfied only at finite times. This fact renders the proof for the second part of the \textquotedblleft no-go\textquotedblright\ theorem invalid.

\section{Discussion and Outlook\label{section:discussion}}

\subsection{Summary.}

In this paper, we considered the model of instanton crystal, i.e., spontaneously broken state characterized by a periodic structure of the order parameter in imaginary time.
Staying in the context of the mean-field approximation, we have focused on the dynamical autocorrelation function $C^\BT(t)$ of the order parameter.
In the model we studied, the said function describes the dynamical correlations of a macroscopic arrangement of loop currents.
As such, the relevant correlation function can be, in principle, accessed by means of magnetic neutron scattering which should be sensitive to the magnetic fields induced by the loop currents.

In order to compute the dynamical correlation function, we have performed the analytical continuation from imaginary to real times. As we have shown, the typical procedure based on the fourier representation of correlation functions is ambiguous in this case. Instead, we proposed the method of continuation directly in time domain, which is based on analytical continuation of a system of differential equations satisfied by the mean-field configuration.
The important feature of the method is that it is straightforward to implement numerically.

The results of our calculations indicate that $C^\BT(t)$ exhibits non-trivial slowly decaying oscillations.
Peculiar feature of these oscillations is that their amplitude is periodically modulated, moreover, the period of the modulation corresponds to the bosonic mode frequency. Unfortunately, these results are not enough to make a definitive conclusion about the behaviour of $C^\BT(t)$ in the limit $t\rightarrow+0$. One possibility is that the decay saturates at some point and the oscillations survive at long times.
Another possibility, which is rather plausible, is that the oscillations die out, however the correlation function saturates at some non-zero constant.
In this scenario, an instanton crystal in imaginary time would behave like a prethermal time crystal in real time.

\subsection{Discussion}

All the considerations in the paper are rather theoretical and highly involved. To somewhat remedy that,
we suggest the following qualitative picture of what is going on.

Our picture follows from the initial loop currents statical picture. The
system in the ordered phase can be either in the $+1$ and $-1$ states
(depending on the direction of the loop currents) forming classical
macrosopic bits. The phase of the instanton crystal is characterized by an
imaginary-time-dependent order parameter $b\left( \tau \right) $. We
interpret its oscillating behavior as a possibility for the system to be in
both $+1$ and $-1$ states simultaneously forming a state
\begin{equation}
\left\vert \psi \right\rangle =\alpha \left\vert 1\right\rangle +\beta
\left\vert -1\right\rangle  \label{n8}
\end{equation}%
with real $\alpha $ and $\beta $ satisfying the normalization condition $%
\alpha ^{2}+\beta ^2=1$. In the language of the order parameter $b\left(
\tau \right) $ the degeneracy with respect to $\alpha $ and $\beta $
corresponds to an arbitrary position $\tau _{0}$ of the
instanton-antiinstanton lattice. In the limit of a large period of the
lattice, the probability of the finding system either in the state $%
\left\vert 1\right\rangle $ or in state $\left\vert -1\right\rangle $
is close to unity but, generally, it is a superposition of the both.

The period of the oscillations can rather
easily be extracted from the picture presented here. In the case $%
k\rightarrow 1$ corresponding to the large period $4K\left( k\right) /\gamma
$ of the instanton lattice, the gap in the spectrum $2\gamma $ in the limit $%
k\rightarrow 1$ should produce oscillations in real time with the period $%
\pi \gamma ^{-1},$ which is standard for a 2-level system. Our result
obtained in this limit is $2K\left( k^{\prime }\right) /\gamma .$ Using the
limiting value $K\left( k^{\prime }\rightarrow 0\right) =\pi /2$ we obtain
the same period $\pi \gamma ^{-1}$ .

So, the properties of the real-time behavior of the instanton crystal
correlate with those of a macroscopic qubit. The slowness of oscillation decay corresponds to the suppression of the decoherence. It is a collective effect
originating from the special long-range order present in the instanton crystal.

\subsection{Outlook.}

Since all the results have been obtained using the mean-field approximation, the important open question is: what is the influence of the fluctuations around the mean-field configuration?

As we have argued in~\cite{se}, the fluctuations should not destroy the instanton crystal state:
We have considered a system with at least two spatial dimensions, and the imaginary time acts as an additional dimension as well. At the same time, the order parameter in the absence of insttantons corresponds to discrete $\mathbb{Z}_2$-symmetry breaking.

However, the form of the dynamical correlation function may be influenced by the inclusion of fluctuations.
We believe that the important role in this regards may play specifically fluctuations in imaginary time.
As we have mentioned, in the absence of non-local interaction, the instanton configuration is a saddle point: there are fluctuation modes with negative energy. In the limit of large period of instanton lattice, these modes correspond to the displacement of instantons and antiinstantons relative to each other. The introduction of non-local interaction stabilizes the instanton configuration and brings the energies of these ``displacement'' modes above zero.
Nevertheless, we expect these modes to play the role of the important low-energy excitations.

In conclusion, we should say that we are hoping to touch the subject of the fluctuations in the future work.

\begin{acknowledgments}
Financial support of Deutsche Forschungsgemeinschaft (Projekt~EF~11/10-1) is
greatly appreciated.
\end{acknowledgments}

\appendix

\section{Magnetic neutron scattering and correlation function $C^\BT(t).$\label{section:scattering}}

In this section we show that the correlation function $C^{\mathcal{T}}(t)$
computed previously can in principle be directly measured. The model for the
instanton crystal considered here has been suggested for interacting loop
currents and a boson mode that both oscillating in space with the vector $%
\mathbf{Q}_{AF}$. Studying experimentally inelastic magnetic neutron
scattering might be a proper tool to measure such magnetic structures and we
show now that the correlation function $C^{\mathcal{T}}(t)$ is just what is
needed to obtain the scattering cross-section.

Considering neutrons scattered by the instanton crystal we start with a
Hamiltonian $\mathcal{H}$ containing both the neutrons and the instanton
crystal. We write it in the form
\begin{equation}
\mathcal{H}=\hat{H}_{n}+\hat{H}+\hat{V}_\mathrm{int}.  \label{nc1}
\end{equation}%
In Eq. (\ref{nc1}), $\hat{H}_{n}$ is the Hamiltonian of free neutrons%
\begin{equation}
\hat{H}_{nt}=\sum_{p}E_{\mathbf{p}}d_{p}^{+}d_{p},  \label{nc2}
\end{equation}%
where $E_{\mathbf{p}}=\mathbf{p}^{2}/2M$ is the kinetic energy, and $%
d_{p}^{+}$, $d_{p},$ are creation and destruction operators of neutrons. The
Hamiltonian $\hat{H}$ describes the system of interacting electrons and
bosonic mode studied here, and $\hat{V}_{int}$ stands for the interaction of
neutron magnetic moments with the magnetic field created by electron
currents. At the moment, we write this term without going into details of
the interaction in the form%
\begin{equation}
\hat{V}_{int}=\Gamma \sum_{\mathbf{q}}d_{\mathbf{q}}^{+}d_{\mathbf{q+Q}_{AF}}%
\hat{A}_0,  \label{nc3}
\end{equation}%
where operator $\hat{A}_0$ has been introduced in Eq.~\eqref{dynamical_cf}.

We assume that the interaction constant $\Gamma $ is small and use for
calculation of transition rate the perturbation theory in $\Gamma$.

The initial state of neutron and system at $t\rightarrow -\infty$ is given by a function $\Psi_0 = |\bp\rangle\otimes|\Phi_i\rangle$,
where $\mathbf{p}$ is the initial momentum of neutron, and $\Phi_i$ is the initial state of the instanton crystal.
We write the Schr\"odinger equation for the function $|\Psi(t)\rangle$ in the interaction representation as
\begin{equation}
 i\frac{d|\Psi(t)\rangle}{dt} = \hat{V}_\mathrm{int}(t) |\Psi(t)\rangle, \label{ir-schr},
\end{equation}
where
\begin{equation}
 \hat{V}_\mathrm{int} (t) =
 \Gamma e^{\lambda t}\sum_\bq e ^{i\hat{H}_n t}d_{\mathbf{q}}^{+}d_{\mathbf{q+Q}_{AF}} e^{-i\hat{H}_n t}\hat{A}_0(t)
\end{equation}
The parameter $\lambda $ is infinitesimally small, $\lambda \rightarrow +0$
and is, as usual, introduced as adiabatic switching of the interaction at $%
t=-\infty $. Time-dependent operator of loop currents $\hat{A}_0(t)$ has been defined in Eq.~\eqref{time-operator}.

Equation~\eqref{ir-schr} is solved representing $|\Psi(t)\rangle$ as
\begin{multline}
 |\Psi(t)\rangle = \BT e^{-i\int_{-\infty}^t dt^\prime \hat{V}_\mathrm{int}(t^\prime)}|\Psi_0\rangle
 \approx \\ \approx
 \left[1 - i\int_{-\infty}^t dt^\prime \hat{V}_\mathrm{int}(t^\prime)\right] |\Psi_0\rangle.
\end{multline}

We are interested in a matrix element $a_{fi}$ of the transition between
the initial state $i$ and the final state $f$ of the form $|\Psi_f\rangle = |\bp^\prime\rangle\otimes |\Phi_f\rangle$ and we write
\begin{equation}
 a_{fi} = -i\Gamma \int_{-\infty}^t dt^\prime e^{\lambda t^\prime} e^{i(E_{\bp^\prime}- E_\bp)t^\prime} \langle \Phi_f|\hat{A}_0(t^\prime)|\Phi_i\rangle.
\end{equation}
Then, we obtain
\begin{multline}
 |a_{fi}|^2 = \Gamma^2 \int_{-\infty}^t dt^\prime e^{\lambda t^\prime} \int_{-\infty}^t dt^{\prime\prime}
 e^{\lambda t^{\prime\prime}} e^{-i(E_{\bp^\prime} - E_\bp)(t^\prime-t^{\prime\prime})}
 \times \\ \times 
 \langle \Phi_i|\hat{A}_0(t^\prime)|\Phi_f\rangle \langle \Phi_f|\hat{A}_0(t^{\prime\prime})|\Phi_i\rangle.
\end{multline}

Up until now, we considered particular initial and finite states of instanton crystal.
In reality, we need to sum over unobserved final states $|\Phi_f\rangle$
and perform thermal averaging over initial states $|\Phi_i\rangle$, which leads us to the following expression
\begin{multline}
 \overline{|a_{fi}|^2} = \lim_{\lambda\rightarrow+0}\frac{\Gamma^2}{\BZ}\iint\limits_{-\infty}^t dt^\prime dt^{\prime\prime} e^{\lambda (t^\prime+t^{\prime\prime})} 
 \\
 e^{-i(E_{\bp^\prime} - E_\bp)(t^\prime-t^{\prime\prime})}
 \ttr{e^{-\beta(\hat{H} - \mu\hat{N})} \hat{A}_0(t^\prime) \hat{A}_0(t^{\prime\prime})}
 = \\ =
 \lim_{\lambda\rightarrow+0} \Gamma^2V \iint\limits_{-\infty}^t dt^\prime dt^{\prime\prime} e^{\lambda(t^\prime + t^{\prime\prime})}\\ e^{-i(E_{\bp^\prime} - E_\bp)(t^\prime-t^{\prime\prime})} C^>(t^\prime - t^{\prime\prime})
\end{multline}

Now we want to take the limit $t\rightarrow +\infty$ and calculate the
transition rate. It is convenient to start with the Fourier
transform $C^>(\omega)$ of the function $C^>(t^\prime-t^{\prime\prime})$:
\begin{equation}
 C^>(t^\prime-t^{\prime\prime}) = \int_{-\infty}^{+\infty}\frac{d\omega}{2\pi}C^\omega(\omega) e^{-i\omega(t^\prime-t^{\prime\prime})},
\end{equation}
where
\begin{equation}
 C^>(\omega)=\int_{-\infty}^{\infty}dt e^{i\omega t} C^>(t).
\end{equation}
Then, we have
\begin{multline}
\lim_{\lambda \rightarrow +0}\iint\limits_{-\infty }^{t}dt_{1}dt_{2} e^{\lambda \left( t_{1}+t_{2}\right) }e^{i\left( E_{\mathbf{p}}-E_{%
\mathbf{p}^{\prime }}\right) t}C^{\mathcal{T}}(t_{1}-t_{2})
=\\ =
\frac{1}{2\pi }\lim_{\lambda \rightarrow +0}\int\limits_{-\infty }^{\infty
}d\omega\iint\limits_{-\infty }^{t}dt_1 dt_2 e^{\lambda \left( t_{1}+t_{2}\right)
}e^{-i\omega \left( t_{1}-t_{2}\right) }  \times \\
\times e^{i\left( E_{\mathbf{p}}-E_{\mathbf{p}^{\prime }}\right) \left(
t_{1}-t_{2}\right) }C^{\mathcal{T}}(\omega )  = \\
=\frac{1}{2\pi }\lim_{\lambda \rightarrow +0}\int_{-\infty }^{\infty }d\omega%
\frac{2\lambda e^{2\lambda t}}{\lambda ^{2}+\left( \omega -E_{\mathbf{p}}+E_{%
\mathbf{p}^{\prime }}\right) ^{2}}C^{\mathcal{T}}(\omega ) ,
\label{nc11a}
\end{multline}%
and, finally,
\begin{multline}
\frac{d}{dt} \overline{|a_{fi}|^2} = 
 \\
=\frac{\Gamma ^{2} V}{2\pi }\int_{-\infty }^{\infty }d\omega C^>(\omega
)\lim_{\lambda \rightarrow +0}\frac{2\lambda }{\lambda ^{2}+\left( \omega
-E_{\mathbf{p}}+E_{\mathbf{p}^{\prime }}\right) ^{2}}  = \\
=\Gamma ^{2}V\int_{-\infty }^{\infty }d\omega C^>(\omega )\delta \left(
\omega -E_{\mathbf{p}}+E_{\mathbf{p}^{\prime }}\right)   = \\
=\Gamma ^{2}VC^>(E_{\mathbf{p}}-E_{\mathbf{p}^{\prime }})
\notag
\end{multline}%

This is actually the standard way of derivation of the scattering
cross-section in quantum mechanics. One can conclude that the magnetic
neutron scattering cross-section is directly given by the function
$C^>(E_{\mathbf{p}}-E_{\mathbf{p}^{\prime }})$ and can be written as
\begin{equation}
\frac{d^{2}\sigma \left( \mathbf{q,}E\right) }{dEd\Omega }\propto \Gamma
^{2}VC^>(E)\delta \left( \mathbf{q-Q}_{AF}\right)  \label{nc13}
\end{equation}%
where $E$ is transferred energy and $\mathbf{q}$ transferred momentum.
In its turn, $C^>(t)$ is directly related to the function $C^\BT(t)$, as we have discussed in section~\ref{section:continuation}.

\section{Numerical scheme for analytic continuation\label{section:scheme}}

For an Instanton Crystal consisting of $m$ instanton-antiinstanton pairs, we denote the period of the pair as $W = \beta/m$.
The numerical scheme that we developed in our previous paper~\cite{se} can be used to compute $b(\tau_i)$ and $\vec C_{\bp}(\tau_i)$ for a discrete set of $N$ equidistant points $\tau_i$ belonging to a single period: $0\leq \tau_i< W$. Consequently, solving the system of Eqs.~\eqref{pseudospins_rt} and~\eqref{gap:diff_final_rt} for each of $\tau_i$, one obtains the values of $b(\tau)$ and $\vec S_\bp(\tau)$ at points $\tau_i+it_j$. With this knowledge, the correlation function can be approximated as
\begin{equation}
 C^\BT(t_j) = \frac{4}{N} \sum_i S_2(\tau_i+it_j) S_2(\tau_i),
\end{equation}
which is the discretized version of Eq.~\eqref{corfunc_mf_continued}. Note also that we used the periodicity of the Instanton Crystal to reduce the sum to a single period.

The important technical detail which deserves to be mentioned is the way the initial conditions for $db(\tau+i0)/dt$ are computed. Using the periodicity in the imaginary time, we rewrite Eq.~\eqref{derivative_init} as
\begin{multline}
 \left.\frac{db(\tau + it)}{dt}\right|_{t=0} = i\frac{U_2}{U_0}\lint_0^W d\tau^\prime\frac{d\tilde K_0(\tau-\tau^\prime| \tilde\omega_0)}{d\tau} b(\tau^\prime)
 +\\ \* +
 2(U_0 + U_2) \int\frac{d\bp}{(2\pi)^2}\varepsilon_\bp^-S_{1\bp}(\tau),\label{derivative_period_reduced}
\end{multline}
where
\begin{multline}
 \tilde K_0(\tau-\tau^\prime|\tilde\omega_0) = \sum_{k = 0}^{m-1} K_0(\tau - \tau^\prime - kW|\tilde\omega_0)
 = \\ \* =
 \frac{\tilde\omega_0\cosh{\left[\tilde\omega_0\left(\frac{W}{2} - |\tau-\tau^\prime|\right)\right]}}{2\sinh{\frac{W\tilde\omega_0}{2}}}.
\end{multline}
Note that the functional form of $\tilde K_0(\tau-\tau^\prime|\tilde\omega_0)$ is identical to the functional form of $K_0(\tau-\tau^\prime|\tilde\omega_0)$. The only difference is that the inverse temperature $\beta$ is replaced with the the Instanton Crystal period $W$.

Correspondingly,
\begin{multline}
 \frac{d\tilde K_0(\tau-\tau^\prime| \tilde\omega_0)}{d\tau} = -\sign{(\tau-\tau^\prime)}
 \times \\ \* \times
 \frac{\tilde\omega_0\sinh{\left[\tilde\omega_0\left(\frac{W}{2} - |\tau-\tau^\prime|\right)\right]}}{2\sinh{\frac{W\tilde\omega_0}{2}}}.
\end{multline}
Finally, we can write the discretized version of Eq.~\eqref{derivative_period_reduced} as
\begin{multline}
  \left.\frac{db(\tau_i + it)}{dt}\right|_{t=0} = i\frac{U_2}{U_0}\frac{W}{N}\sum_j \frac{d\tilde K_0(\tau_i-\tau_j| \tilde\omega_0)}{d\tau} b(\tau_j)
 +\\ \* +
 2(U_0 + U_2) \int\frac{d\bp}{(2\pi)^2}\varepsilon_\bp^-S_{1\bp}(\tau_i),\label{derivative_discretized}
\end{multline}
where it is assumed that $\sign(0) = 0$.

In the absence of non-local interaction, $U_2=0$, functions $b(\tau+it)$ has the poles along the lines $\tau = 0$ and $\tau=W/2$. This poles survive even for finite $U_2$. In order to avoid them, we chose the set of time points $\tau_i$ to be $\tau = W/N(i-1/2)$ for $1\leqslant i\leqslant N$.

As in our previous work, \cite{se}, we used programming language Julia~\cite{Julia-17} to perform the numerical calculations presented in the paper.

\bibliography{ITC}

\end{document}